%% file: voidbao_dr11.tex
\documentclass[apsrev4-1,prd,twocolumn,showpacs]{revtex4-1}
\usepackage{graphicx}
\usepackage{dcolumn}
\usepackage{bm}
\usepackage{natbib}
\usepackage{multirow}
\usepackage{url}
\usepackage{color}
\usepackage{xcolor}
\usepackage{tikz}
\usepackage{float}

\usepackage{breakurl}

\input{HEADER_prl}

\begin{document}

\title[]{Signatures of the Primordial Universe from Its Emptiness:\\Measurement of Baryon Acoustic Oscillations from Minima of the Density Field}

\author{Francisco-Shu Kitaura$^{1*}$,  Chia-Hsun Chuang$^{1}$, Yu Liang$^{2}$, Cheng Zhao$^{2}$, Charling Tao$^{2,3}$, Sergio Rodr{\'i}guez-Torres$^{4,5,6}$, Daniel J.~Eisenstein$^{7}$, H{\'e}ctor Gil-Mar{\'i}n$^{8,9}$,  Jean-Paul Kneib$^{10,11}$, Cameron McBride$^{7}$, Will J.~Percival$^{12}$, Ashley J.~Ross$^{12,13}$, Ariel G.~S{\'a}nchez$^{14}$, Jeremy Tinker$^{15}$, Rita Tojeiro$^{16}$, Mariana Vargas-Magana$^{17}$, Gong-Bo Zhao$^{18,12}$
}   
\medskip
\affiliation{$^{1}$Leibniz-Institut f\"ur Astrophysik Potsdam (AIP), An der Sternwarte 16, D-14482 Potsdam, Germany\\
$^{2}$Tsinghua Center of Astrophysics and Department of Physics, Tsinghua University, Beijing 100084, China.\\
$^{3}$Aix-Marseille Universit\'{e}, CNRS/IN2P3, CPPM UMR 7346, 13288 Marseille, France\\
$^{4}$Instituto de F\'isica Te\'orica, (UAM/CSIC), Universidad Aut\'onoma de Madrid, Cantoblanco, E-28049 Madrid, Spain\\
$^{5}$Campus of International Excellence UAM+CSIC, Cantoblanco, E-28049 Madrid, Spain\\
$^{6}$Departamento de F\iısica Te\'orica M8, Universidad Autonoma de Madrid (UAM), Cantoblanco, E-28049, Madrid, Spain\\
$^{7}$ Harvard-Smithsonian Center for Astrophysics, 60 Garden Street, Cambridge, Massachusetts 02138, USA\\
$^8$ Sorbonne Universités, Institut Lagrange de Paris (ILP), 98 bis Boulevard Arago, 75014 Paris, France \\
$^9$ Laboratoire de Physique Nucléaire et de Hautes Energies, Universit\'e Pierre et Marie Curie, 4 Place Jussieu, Tour 22, 1er étage, 75005 Paris, France \\
$^{10}$ Laboratoire d’Astrophysique, Ecole Polytechnique Fed\'{e}rale de Lausanne, CH-1015 Lausanne, Switzerland\\
$^{11}$ Aix Marseille Universit\'{e}, CNRS, LAM (Laboratoire d’Astrophysique de Marseille) UMR 7326, F-13388, Marseille, France\\
$^{12}$ Institute of Cosmology \& Gravitation, University of Portsmouth, Dennis Sciama Building, Portsmouth PO1 3FX, United Kingdom\\
$^{13}$ Center for Cosmology and AstroParticle Physics, The Ohio State University, Columbus, Ohio 43210, USA\\
$^{14}$ Max-Planck-Institut f\"ur Extraterrestrische Physik, Postfach 1312, Giessenbachstrasse, 85741 Garching, Germany\\
$^{15}$ Center for Cosmology and Particle Physics, New York University, 4 Washington Place, New York 10003, USA\\
$^{16}$ University of St Andrews, North Haugh, St Andrews Fife, KY16 9SS, United Kingdom \\
$^{17}$ Instituto de F\'isica, Universidad Nacional Aut\'onoma de M\'exico, Apdo. Postal 20-364, M\'exico\\
$^{18}$ National Astronomy Observatories, Chinese Academy of Science, Beijing, 100012, People's Republic of China
}
\email{kitaura@aip.de}

\date{\today}

\begin{abstract}
 Sound waves from the primordial fluctuations of the Universe imprinted in the large-scale structure, called baryon acoustic oscillations (BAOs), can be used as standard rulers to measure the scale of the Universe. 
These oscillations have already been detected 
%in the cosmic microwave background, in the Lyman alpha forest, and  
 in the distribution of galaxies. Here we propose to measure BAOs from the troughs (minima) of the density field.
 Based on two sets of accurate mock halo catalogues with and without BAOs in the seed initial conditions, we demonstrate that the BAO signal cannot be obtained from the clustering of classical disjoint voids, but is clearly detected from overlapping voids. The latter represent an estimate of all troughs of the density field.
 We compute them from the empty circumsphere centers constrained by tetrahedra of galaxies using Delaunay triangulation. Our theoretical models based on an unprecedented large set of detailed simulated void catalogues are remarkably well confirmed  by observational data. We use the largest recently publicly available  sample of  luminous red galaxies from SDSS-III BOSS DR11 to unveil for the first time a $>$3$\sigma$ BAO detection from voids in observations. 
  Since voids are nearly isotropically expanding regions, their centers represent the most quiet places in the Universe, keeping in mind the cosmos  origin and providing a new promising window in the analysis of the cosmological large-scale structure from galaxy surveys. 
\end{abstract}

\pacs{98.80.-k, 98.80.Es,98.65.Dx}

\maketitle

\setcounter{footnote}{0}

In the primordial baryon-photon plasma of our Universe, overpressured regions triggered sound waves that stalled at the recombination epoch, imprinting spheres of overdensity fluctuations,  measurable in the matter power spectrum as an oscillatory pattern, the so-called baryon acoustic oscillations (BAOs). Any dark matter tracer should encode this signal in its spatial distribution  either at early or late cosmic times after cosmic evolution  (see Refs.~\citep[][]{PY70,SZ70,BG03,SE05}). In fact these oscillations have already been detected in the cosmic microwave background anisotropies (see Refs.~\citep[][]{WMAP103,WMAP711,WMAP913,PLANCKBAO14}), in the distribution of galaxies (see Refs.~\citep[][]{CPP05,EZH05,PRE10,BKB11,BBC11,AAB14}), and more recently in the distribution of the Lyman alpha forest (see Refs.~\citep[][]{BDR13,SIK13,DBB15}). For a review on BAOs and their cosmological implications, see \citet[][]{ABB14}.

\begin{figure}
\includegraphics[width=8.cm]{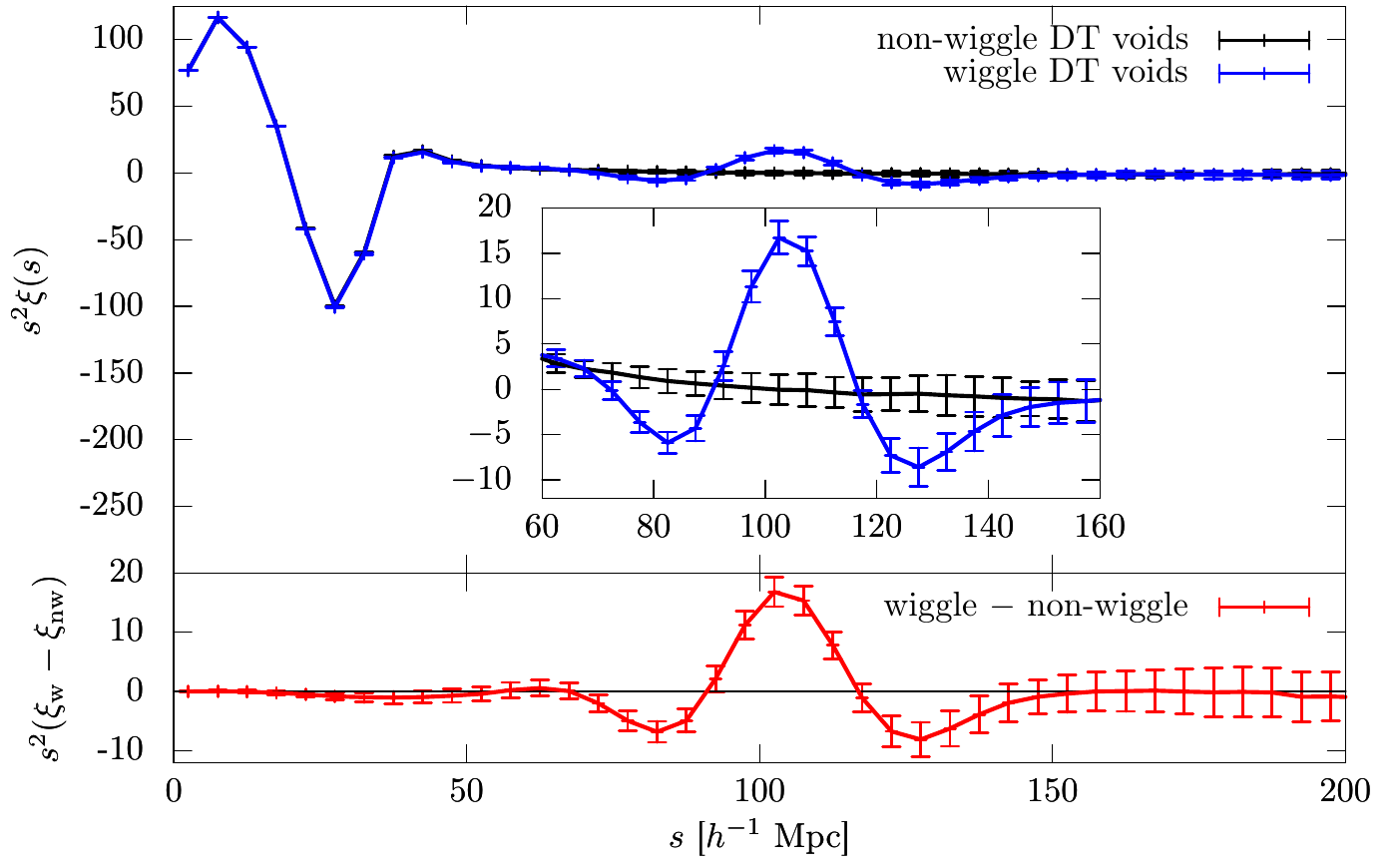}
\caption{\label{fig:sim} Correlation functions  for the set of 100 \textsc{patchy} (full cubic volume at mean redshift 0.56) void tracer mock catalogues (without observational effects) based on seed perturbations with and without BAOs. {\it Upper panel:} Mean and variance for the following cases: (1) with BAOs, blue solid line and blue error bars, respectively; (2) without BAOs (``nonwiggle''): black solid line  and black error bars, respectively. {\it Lower panel:} Corresponding residual (red solid line and red error bars).}
\end{figure}

Their characteristic scale can be used as a standard ruler to measure the evolving scale of the Universe and to constrain the nature of its driving force, the dark energy component. For this reason a large number of surveys have focused on measuring BAOs, or have included them as an integral part of their science, such as the 2dFGRS \citep{CDM01}, the SDSS \citep{SDSS00}, the WiggleZ \citep[][]{wigglez2010}, the BOSS \citep[][]{boss2011}, the SDSS-IV/eBOSS, the DESI/BigBOSS \citep[][]{bigboss2011}, the DES \citep[][]{2006astro.ph..9591A}, the LSST  \citep[][]{lsst2012}, the J-PAS \citep[][]{jpas2014}, the 4MOST \citep[][]{4most}, or the EUCLID survey \citep[][]{euclid2009}.

Ever since the first detection of the giant Bo{\"o}tes void in 1981 \citep[][]{KOS81} 
and with the nascent era of galaxy surveys,  more evidence for the existence of voids has been  found.
The presence of voids in the large-scale structure was considered a manifestation of  cosmological structure formation  transforming the homogeneous Universe into a complex cosmic web structure. This picture was confirmed through numerical simulations, see, e.g., Refs.~\citep[][]{Klypin1983,Blumenthal1984,DEF85}.
 The classification of voids based on galaxy surveys has turned into a common practice, (see, e.g., the CfA \citep[][]{LGH86,VGP94},  the IRAS \citep[][]{EP97}, Las Campanas \citep[][]{MAE00}, the PSCz \citep[][]{PB02}, 
the 2dFRGS \citep[][]{CCG04,HV04,PBP06}, the DEEP2  \citep[][]{CCW05}, the 2MRS  \citep[][]{NKH14}, the SDSS survey \citep[][]{PWJ11,VBT12,PVH12,NH14,SLW14,BKH15}, and the VIMOS survey \citep[][]{2014A&A...570A.106M}). %%\citep[see e.g.~][]{SLW14}.
%\cite[see][]{LGH86,VGP94,EP97,MAE00,PB02,CCG04,HV04,CCW05,PVH12,SLW12,NH14,SLW14}. 
Nevertheless,  voids are usually considered to be very large rare objects, as compared to galaxies. Their probability distribution function can be used to constrain cosmology in an analogous way to galaxy clusters (see Ref.~\citep[][]{BPP09}).
The statistics of voids has been studied for a long time (see, e.g., Refs.~\citep[][]{W79,PP86,B90,EEG91,BL02})
, and  an excursion set formalism analogous to the one describing the formation of halos (the compact collapsed dark matter objects hosting galaxies) has been developed (see Refs.~\citep[][]{SW04,FP06,PLS12,JLH13}). 
Those studies hint towards a hierarchical picture, in which voids can  form merger trees through cosmic evolution (see Ref.~\citep[][]{AWA10}).
Considerable efforts have been made to understand the nature and evolution of voids through theoretical studies with semianalytic studies (see, e.g., Refs.~\citep[][]{MW02,BHT03} 
and simulations see, e.g., Refs.~\citep[][]{BCG92,DCG93,WK93,GLK03,CSD05,PWJ08,ESH11}).
%\cite[e.g.~][]{BCG92,GLK03,CSD05}.
%

Nevertheless, there are many different definitions of voids (see Refs.~\citep[][]{AM98,FP01,GLK03,SW04,CSD05,Netal05,2005EPJB...47...93G,PPH06,PWJ07,SHH12,AHK12,FNS12,CWJ13,2015ApJ...799...95W}), which do not necessarily agree with each other (see, e.g., Ref.~\citep[][]{CPF08}).

From a practical perspective, voids have recently been proposed to give additional cosmological constraints, not only according to their statistics, but also according to their shape. 
The void ellipticity was proposed to probe dark energy (see Refs.~\citep[][]{PL07,LW10,B12,2015arXiv150307690P}),
 and to make the Alcock-Paczy{\'n}ski test \citep[][]{SPW14}.%%%%
In particular, they can be used to test gravity (see, e.g., Refs.~\citep[][]{B12,LCC15,C15}) %%%%
dynamical dark energy \cite{B12}, coupled dark energy \cite{L11},  and modified gravity \cite{MS09,LCC15}. %%%
They can also be used to measure the Sachs Wolfe effect  \cite{PLANCKISW14}. %%%
However, their sparse population and low signal-to-noise ratio have made them less interesting for clustering analysis. Little work can be found on the measurement of the correlation function of voids; see, however, Refs.~\citep[][]{GCW95,padilla05,HSW14} and, in particular, the recent pioneering study on observations (see Ref.~\citep[][]{CJS15}).

\begin{figure}
\includegraphics[width=8.cm]{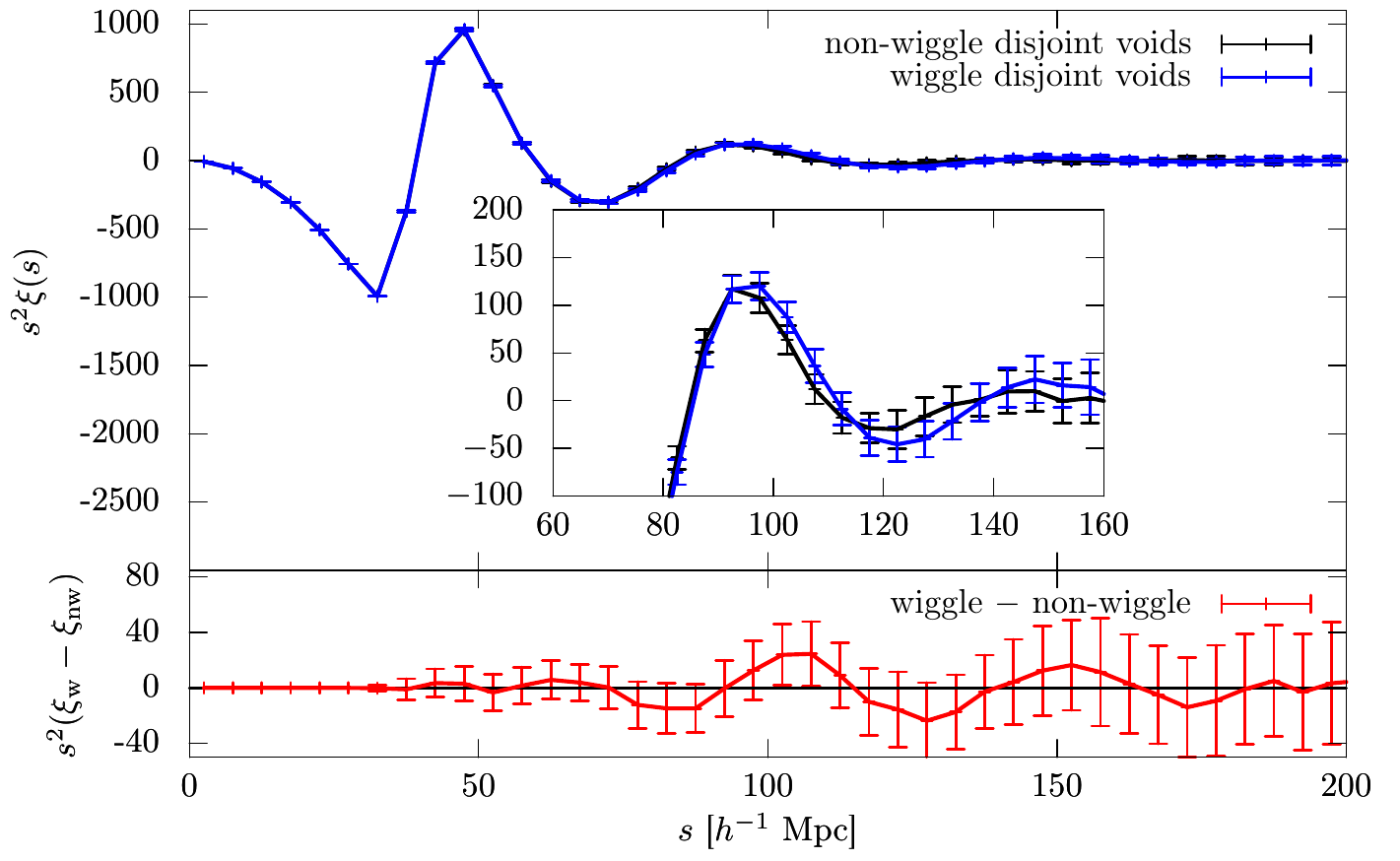}
\caption{\label{fig:nosim} Same as Fig.~\ref{fig:sim} for disjoint voids.}
\end{figure}

In this Letter, we propose for the first time, using the troughs  of the density field (from now on called void tracers), meaning the minima in the overdensity field, to obtain additional measurements of the BAOs from the ones corresponding to galaxies.  
We have developed a Delaunay triangulation void finder (\textsc{dive}) based on empty circumspheres constrained by tetrahedra of galaxies Zhao {\it et~al.} (see companion paper Ref.~\citep[][]{2015arXiv151104299Z}). Our voids are close to the classical definition  as spherical underdense regions (see, e.g., Refs.~\citep[][]{W79,PBP06}),  including, however, as a crucial difference, overlapping spheres, since we are interested in the distribution of troughs of the density field and account, in this way, for the shape of empty regions. 

Our definition crucially increases the statistics of void tracers by about 2 orders of magnitude in contrast to previous studies, in which  voids are treated as large connected regions, that do not overlap at all, or overlap only marginally (see, e.g., Refs.~\citep[][]{PBP06,HSW14,CJS15}).  The speed of the \textsc{dive} void finder has been determinant for this project taking only of the order of minutes to find all the void tracers associated with about half a million objects and with little memory requirements (on a single core: $\sim$18 mins and $\sim$ 5 Gb, respectively).

In Liang {\it et al.} (see companion paper, Ref.~\citep[][]{2015arXiv151104391L}),  we have studied for the first time the BAO signal with this void definition on mock catalogues predicting a characteristic correlation function, which includes dips on scales smaller and larger to the BAO peak. These features were exploited to develop a model-independent  signal-to-noise estimator, used in turn to determine  the  radius cuts that provide the optimal signal-to-noise ratio for the BAO signal.

\begin{figure}
\includegraphics[width=8.cm]{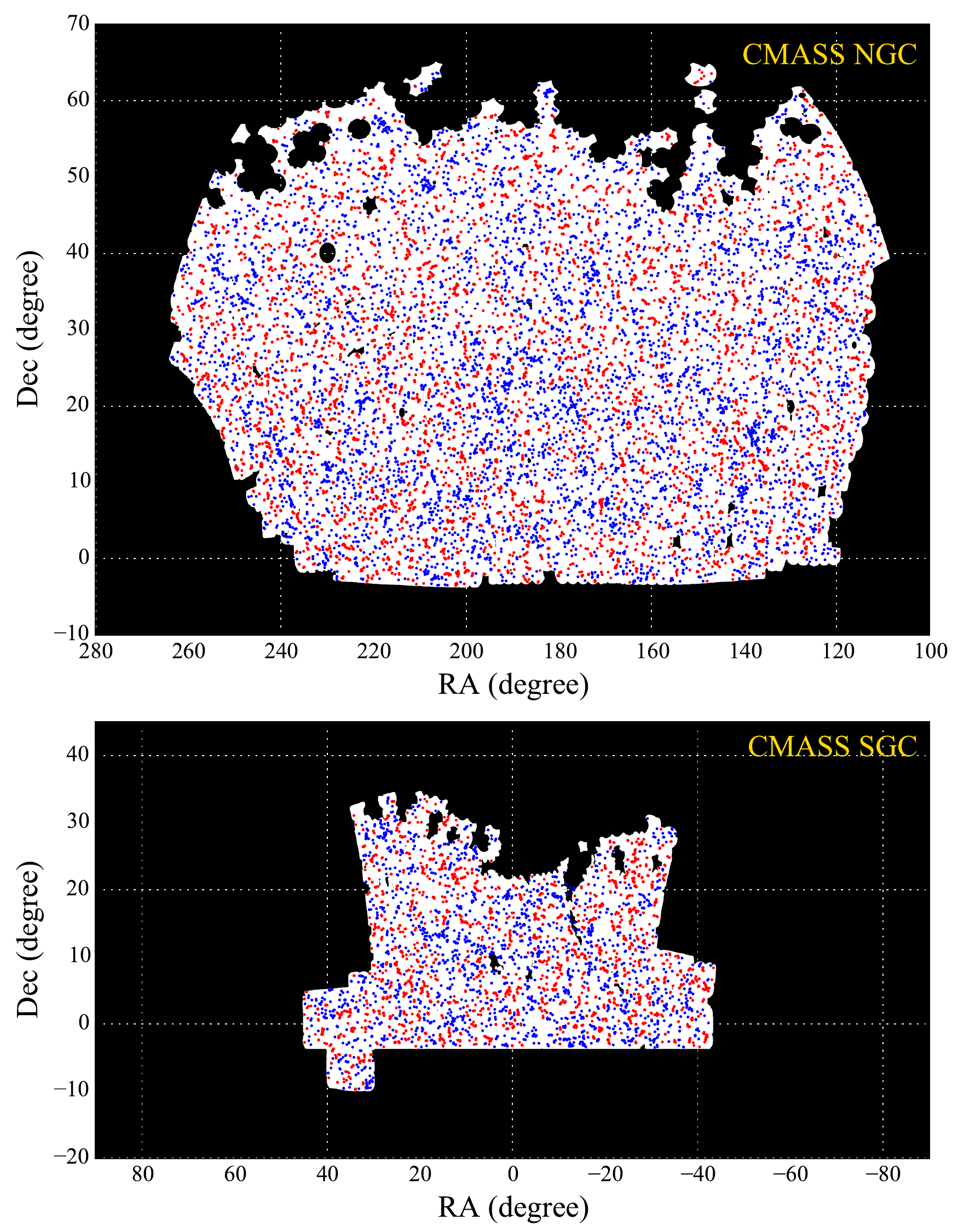}
\caption{\label{fig:sky} Sky projection in right ascension (RA) and declination (DEC) of the BOSS DR11 CMASS LRGs (red symbols) and the corresponding void tracer (blue symbols) catalogues. {\it Upper panel}: Northern galactic cap NGC. {\it Lower panel}: Southern galactic cap (SGC).   Void tracers obtained in unobserved regions or holes in the mask (caused by e.g.~stars) have been accordingly been removed.}
\end{figure}

In this Letter, we aim to extend the signal-to-noise estimator to detect the BAO signal from voids based on observational data.

To this end, first we  define a control sample of accurate mock galaxy catalogues performed with the \textsc{patchy} code (Ref.~\citep[][]{KitauraPatchy}). In particular, we have produced 100 mocks for each of the following cases: catalogues with and without  baryon acoustic oscillations (``wiggle'' and ``nonwiggle'' case, respectively) in the initial conditions used to simulate structure formation. In particular we consider complete samples of halos (main and subhalos) in cubic volumes of (2.5 $h^{-1}$ Gpc)$^3$ with number density 3.5 $10^{-4}$ $h^{3}$ Mpc$^{-3}$, similar to the one of the BOSS CMASS galaxy sample at a mean redshift $z=0.56$. The parameters of the \textsc{patchy} code  have been calibrated with the large BigMultiDark $N$-body simulation (Ref.~\citep[][]{Klypin2014}) to accurately match the two- and the three-point statistics (such parameters can be found in Ref.~\citep[][]{KGS15}). { The cosmological parameters have been consistently chosen to be within $\Lambda$ cold dark matter Planck cosmology with  $\Omega_{\rm M}$ = 0,307115; $\Omega_{\rm b}$ = 0,048206; $\sigma_8$ = 0,8288; $n_{\rm s}$ = 0,9611, and a Hubble constant ($H_0$ = 100 $h$ km\,s$^{-1}$Mpc$^{-1}$) given by $h$ = 0,6777.} 

The accuracy of these catalogues has been further demonstrated in several recent papers (see Refs.~\citep[][]{Zhao15,ChuangComp15}). 

We have run the \textsc{dive} void finder for circumspheres with radii $\geq16$ $h^{-1}$ Mpc on these sets of catalogues in real space, and computed the corresponding correlation functions.
The results do not show any signal in the ``nonwiggle'' case, as expected, while the ``wiggle'' case shows a  significant BAO signal  (see Fig.~\ref{fig:sim}). Hence, both sets of simulations demonstrate that the BAO signal from voids is really present in our mock catalogues, and we confirm the findings in Liang {\it et al.} (see companion paper, Ref.~\citep[][]{2015arXiv151104391L}). 
The two dips around the BAO peak and a singularity around the size (diameter) of the smallest void ($\sim$30 $h^{-1}$ Mpc) due to the void exclusion effect can also be clearly seen in that Fig.~\ref{fig:sim}. Importantly, the BAO peak is not only seen in the residual after extracting the ``nonwiggle'' from the ``wiggle'' mock catalogues (see lower panel in Fig.~\ref{fig:sim}), but directly in the correlation function based on the catalogues containing the BAO signal in the seed perturbations (see upper panel Fig.~\ref{fig:sim}). 
This is not the case when analyzing disjoint voids (see Fig.~\ref{fig:nosim}). The oscillation patterns seen in the correlation functions are not related to the BAOs, but are due to hard sphere exclusion effects when the filling factor is high (see Ref.~\citep[][]{W63}), as they can be found both in the ``wiggle'' and ``nonwiggle'' mock catalogues. There are only tiny differences in the modulation of these oscillations caused by BAOs which can only be found in the residuals with large error bars (compare upper and lower panels in Fig.~\ref{fig:nosim}).

\begin{figure}
\begin{tabular}{c}
\includegraphics[width=8.cm]{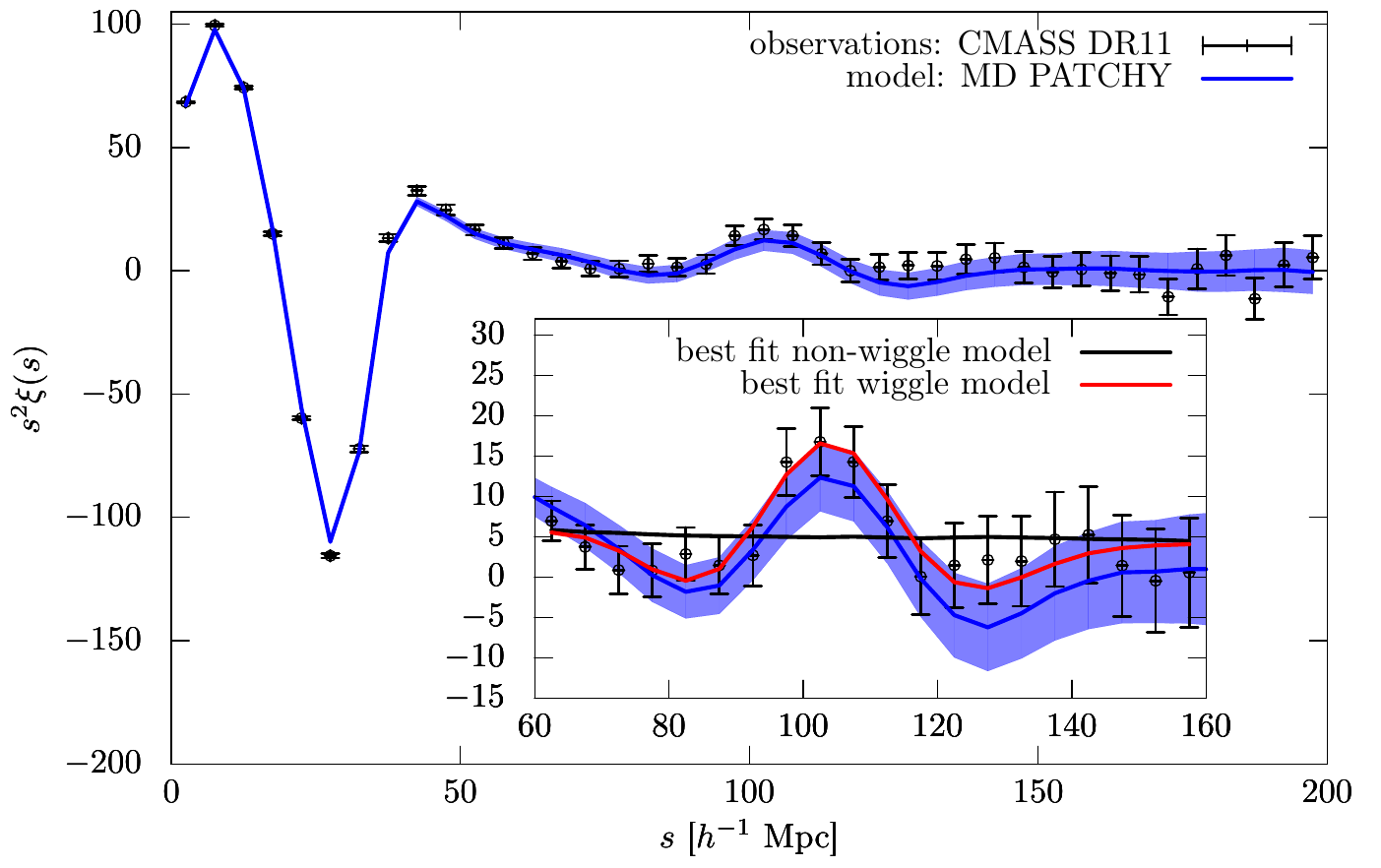}
\put(-135,33){\tiny BAO 3.2 $\sigma$ detection}
\end{tabular}
\caption{\label{fig:data} Correlation functions  for the BOSS DR11 CMASS  void tracer catalogue (black error bars) and the mean (blue) and 1-$\sigma$  region (blue shaded) of the corresponding 1,000 light-cone (including evolution from redshift 0.43 to 0.7) \textsc{MultiDark patchy} DR11 CMASS  mock void catalogues (including observational effects: survey geometry, mask, radial selection function, and redshift-space distortions). The ``wiggle'' and ``nonwiggle'' best fitting models are represented by the red and black solid lines, respectively.}
\end{figure}

We have verified that the majority of the void tracers considered  are located in expanding regions and that they are anticorrelated to the halos, hereby demonstrating that our definition of voids yields additional tracers of the large-scale structure  (see Ref.~\citep[][]{2015arXiv151104299Z}).

To detect the void tracer BAO signature in observations, we need to consider mocks resembling the BOSS DR11 CMASS sample in our analysis, including survey geometry, radial selection effects, bias evolution and redshift space distortions (RSDs). 

This work uses data from the Data Release DR11 \citep[][]{Alam15} of the Baryon Oscillation
Spectroscopic Survey (BOSS) \citep[][]{2011AJ....142...72E}. The BOSS survey
uses the SDSS 2.5 meter telescope at Apache Point Observatory \citep[][]{2006AJ....131.2332G} and the spectra are obtained using the double-armed BOSS
spectrograph \citep[][]{2013AJ....146...32S}. The data are then reduced using the
algorithms described in Ref.~\citep[][]{2012AJ....144..144B}.  The target selection of
the CMASS and LOWZ samples, together with the algorithms used to
create large scale structure catalogues (the \textsc{mksample} code), are
presented in  \citet[][]{2015arXiv150906529R}.

We compute the voids (with radii $\geq16$ $h^{-1}$ Mpc) and the corresponding correlation functions for 1,000 BOSS DR11 CMASS \textsc{MultiDark patchy} mocks (see Ref.~\citep[][]{2015arXiv150906400K}). These galaxy mocks have been calibrated with $N$-body based reference catalogues from the BigMultiDark simulation (see Ref.~\citep[][]{2015arXiv150906404R}) and made publicly available\footnote{\burl{http://data.sdss.org/datamodel/files/BOSS_LSS_REDUX/dr11_patchy_mocks/}}. The radius cut was determined to provide the optimal signal-to-noise ratio for the BAO signal  (see Ref.~\citep[][]{2015arXiv151104391L}).

We follow the methodology presented in Liang {\it et al.} (companion paper, Ref.~\citep[][]{2015arXiv151104391L}) to deal with the survey geometry and radial selection function.
In particular, we use the angular mask from the DR11 galaxy catalogue to filter out the voids identified outside the survey area to construct the observed DR11 void catalogue and the corresponding set of synthetic BOSS DR11 CMASS  \textsc{MultiDark patchy}  void light-cone catalogues. To compute the two-point correlation functions, we need to construct a random void catalogue with the same  geometry (in both angular and radius directions) as the BOSS DR11 CMASS  data. To that purpose we combine 50 BOSS DR11 CMASS  \textsc{MultiDark patchy} void catalogues and reassign the redshift randomly picked from observed data (a.k.a.~shuffle method, e.g., see Ref.~\citep[][]{AAB14}). This procedure will produce  random void catalogues with geometry consistent with the observed data. We avoid using the random galaxy catalogue for the random void catalogue, since the distribution of the voids is different, especially at the boundaries of the survey.

{ Our analysis relies on a factor 2--2.5 more troughs than galaxies  (for CMASS North: 1,212,393 troughs--voids with radii $\geq16$ $h^{-1}$ Mpc--vs 566,940 galaxies; and for CMASS South: 472,868 troughs vs 188,582 galaxies). As an example for the CMASS North we would only have 48,000 disjoint voids.}

Finally, we take the BOSS DR11  data and apply the same analysis algorithms, using the same settings.
A plot  of the sky projection of the galaxies and their corresponding void tracers clearly illustrates how these tracers trace different regions of the cosmic web  (see Fig.~\ref{fig:sky}).
The result of these computations shows a remarkable agreement between the theoretical prediction and the observations even towards large scales in contrast to galaxies (see Fig.~\ref{fig:data}). 
Here we use the ``wiggle'' and ``nonwiggle'' simulations to construct the templates of the fitting models to estimate the significance of the BAO detection. 

We make a cubic spline fit from the ``wiggle'' and ``nonwiggle'' \textsc{patchy} mock correlation function,  $\xi_{\rm w}(s)$ and $\xi_{\rm nw}(s)$, respectively, with $s$ being the separation between two void tracers based on the galaxy distribution in redshift space. These two functions are the basis to construct the ``wiggle''  and ``nonwiggle'' models for determining the BAO significance.
In particular, we apply the following models in the fitting range  $60 < r < 160$ $h^{-1}$ Mpc.
First, we show  a ``wiggle'' model: \be \label{eq:wiggle}
\xi_{\rm th}(s) = A\,[\xi_{\rm w}(s/\alpha) - \xi_{\rm nw}(s/\alpha)] + \xi_{\rm nw}(s/\alpha) + a_0 + a_1/s + a_2/s^2\,,\ee
where $\alpha$ is the rescaling factor of BAO, $A$ is the BAO damping factor, and the polynomial models the systematics for the overall shape following Anderson {\it et al.}~\citep[][]{AAB14}.
And second a ``nonwiggle'' model: \be \xi_{\rm th}(s) = \xi_{\rm nw}(s/\alpha) + a_0 + a_1/s + a_2/s^2\,,\ee which can be obtained from setting $A=0$ in the ``wiggle'' model Eq.~\ref{eq:wiggle}.

{ 
As in \citet[][]{AAB14}, we use a template with fixed cosmology. The measurement of alpha can be interpreted as the ratio between the spherically averaged distance scale $D_{\rm V}(z) \equiv [cz\,(1+z)\,2\,D_A(z)2H^{-1}(z)]^{1/3}$ to the pivot redshift ($z=0.57$)  and the sound horizon scale rs at drag epoch with respect to the fiducial model: $\alpha=[D_{\rm V}/r_{\rm s}]/[D_{\rm V}/r_{\rm s}]_{\rm fid}$,
where $D_A(z)$ is the angular diameter distance and $H(z)$ is the Hubble parameter. 
In general, a theoretical correlation function model should be constructed with parameters \{$\Omega_{\rm M} \, h^2$, $n_{\rm s}$, $\Omega_{\rm b}\, h^2$, $\alpha$\}, where $\alpha$ absorbs the information of dark energy and curvature. In practice, one might ignore the uncertainties of $n_{\rm s}$ and $\Omega_{\rm b} \,h^2$ since they are tightly constrained by CMB. While fixing $\Omega_{\rm M} \,h^2$, we can only measure some quantity which is insensitive to $\Omega_{\rm M}\, h^2$. Therefore, $\alpha$ should be interpreted as $D_{\rm V}/r_{\rm s}$ which is uncorrelated to $\Omega_{\rm M} \,h^2$ (e.g. see Table 2 in Ref.~\citep[][]{Chuang12}).

The significance of the detection was computed from the difference of the best ``wiggle'' and ``nonwiggle'' fits yielding a chi-squared  per degrees of freedom of $\chi^2/{\rm dof}=9.9/15$ for the ``wiggle'' model, $\chi^2/{\rm dof}=20.1/16$ for the ``nonwiggle'' model.
In particular, we measured $\alpha$ by marginalizing over the amplitude $A$, obtaining $\alpha=1.000 \pm 0.022$.
 Converting this finding to an effective distance at $z=0.57$ would correspond to $2057 \pm 45$ Mpc, which is compatible with the finding from galaxies alone (see Ref.~\citep[][]{AAB14}, which found $2056 \pm 20$ Mpc). 
One should note that the chi-squared distribution is not very Gaussian for voids. We would therefore take this measurement as a first-order estimate and work on more robust  measurements in forthcoming  papers. }

 Relying on these models we find a BAO detection with a significance of 3.2 $\sigma$ (see Fig.~\ref{fig:data}). We have used the covariance matrices derived from the set of 1,000 mocks to do this analysis analogously to Anderson {\it et al.} \citep[][]{AAB14}.
As a first approximation we assume in the ``wiggle'' and ``nonwiggle'' models that RSDs can be modeled by a damping term. We plan to investigate RSDs in detail in future work. 
Incompleteness, veto mask, and the fiber collision are taken into account in the DR11 CMASS mock catalogues, and, accordingly, in the void catalogue computations. 
We do not see in the CMASS void correlation function any strong systematic effects, i.e. strong deviations in the correlation function towards large scales, as it  was seen with the CMASS galaxy correlation function (see Refs.~\citep[][]{Ross2012,Chuang2013d}). 
The correlation function behaves very much like the theoretical correlation function from the light-cone mocks.
With the optimal radius cut used in this study we found that the number density of voids is insensitive to the number density of galaxies (see Fig.~4 in Ref.~\citep[][]{2015arXiv151104299Z}). This would explain, why a varying number density of galaxies caused by stellar density systematics does not have a significant impact on the void density across the sky.

{
Question arise when we measure the clustering of voids: What is the information gain from void tracers directly computed from the distribution of galaxies? And how covariant are these tracers  to the galaxies themselves?
The construction of void troughs follows the intuitive physical picture of filling the gaps complementary to the high density peaks occupied by the galaxies. Luminous red galaxies (LRGs) are known to reside in high density regions (see, e.g., Ref.~\citep[][]{KGS15}). We are thus, extending the information on the density fluctuations ($\delta=\rho/\bar{\rho}-1$) to underdense regions ($\delta<0$), which based on this galaxy distribution are otherwise set to a constant value ($\delta=-1$). { Less massive objects, such as emission line galaxies,  could also be used to define underdense regions, but
an extended definition with some stellar mass threshold may be required for the estimation of troughs.}
We note, that small voids are equivalent to groups of quartets of galaxies residing in high density regions (see Ref.~\citep[][]{2015arXiv151104299Z}), and, hence, are expected to deliver redundant information to the galaxies themselves.
This is not the case for the large voids considered in this study. 
In fact, it is clear, that the Delaunay voids we construct from tetrahedra of galaxies encode higher order statistics, further constrained by imposing the circumspheres to be empty, which strongly depends on gravitational evolution of the morphology of the cosmic web and hence, on all the $n$-point statistics of the density field (in particular the three-point statistics, see Ref.~\citep[][]{Frieman1994}). Moreover, our prior knowledge on the radius cut selecting empty circumspheres located in expanding void regions, based on tidal field computations of the underlying dark matter field in simulations (see Zhao {\it et~al.} companion paper \citep[][]{2015arXiv151104299Z}), implicitly incorporates knowledge on the void regions beyond  the one present in the galaxy distribution.
By analyzing the clustering of the troughs (constructed upon the galaxies) we are including higher order information (see \citep[][]{W79}),  potentially circumventing a more complicated mathematical formalism needed to extract the full information encoded in the three-dimensional distribution of galaxies.
{ This is supported by recent theoretical work, demonstrating that most of the information gained in BAO reconstruction comes from the three-point statistics with some contributions from the four-point statistics (see Ref.~\citep[][]{2015PhRvD..92l3522S}), and depends on the environment (see Ref.~\citep[][]{2015PhRvD..92h3523A}). In fact a recent work has presented a 2.8 $\sigma$ detection of BAOs from the three-point correlation function based on BOSS DR12 (see Ref.~\citep[][]{2015arXiv151202231S}). } 
The actual information gain we can get from combining void tracers with galaxies in a multitracer analysis remains to be investigated, including whether voids will improve the cosmological constraints from galaxy clustering alone. This analysis may yield little added value in the presence of data covering the underdense cosmic density field, with, e.g., considerably higher number densities, than that provided by LRGs. 
Nevertheless,  since  void tracers are expected to be less affected by gravitational pull,  BAO reconstruction techniques (see Ref.~\citep[][]{ESS07}) could be less necessary for these tracers, and they may thus yield a less cosmology-dependent estimate of the linear correlation function. We will investigate this in future work. 

}

%In summary, we believe that voids, considered as tracers of the large-scale structure can open a new window to the analysis of the cosmological large-scale structure.

\section*{Acknowledgments}
The authors thank { Michael Wood-Vasey, Christian Wagner, Marcos Pellejero Iba{\~n}ez, Juan E. Betancort-Rijo, Anatoly Klypin, Gustavo Yepes, and Francisco Prada}  for useful discussions.
FSK thanks support from the Leibniz Society for the Karl-Schwarzschild fellowship. 
YL, CT, CZ acknowledge support by Tsinghua University with a 985 grant,
973 programme 2013CB834906, NSFC grants No.~11033003, 11173017, sino french CNRS-CAS international laboratories LIA Origins, and FCPPL.
RT thanks support from the STFC Ernest Rutherford Fellowship.
HGM thanks support from the Labex ILP (reference ANR-10-LABX-63) part of the Idex SUPER, receiving financial state aid managed by the Agence Nationale de la Recherche, as part of the programme Investissements d'avenir under the reference ANR-11-IDEX-0004-02. GZ is supported by the Strategic Priority Research Program of the Chinese Academy of Sciences Grant No. XDB09000000.
The authors also thank the access to computing facilities at Barcelona (MareNostrum),  LRZ (Supermuc), AIP (erebos),  CCIN2P3 (Quentin Le Boulc'h), and  Tsinghua University. 

Funding for SDSS-III has been provided by the Alfred P. Sloan Foundation, the Participating Institutions, the National Science Foundation, and the U.S. Department of Energy Office of Science. The SDSS-III web site is http://www.sdss3.org/.

SDSS-III is managed by the Astrophysical Research Consortium for the Participating Institutions of the SDSS-III Collaboration including the University of Arizona, the Brazilian Participation Group, Brookhaven National Laboratory, Carnegie Mellon University, University of Florida, the French Participation Group, the German Participation Group, Harvard University, the Instituto de Astrofisica de Canarias, the Michigan State/Notre Dame/JINA Participation Group, Johns Hopkins University, Lawrence Berkeley National Laboratory, Max Planck Institute for Astrophysics, Max Planck Institute for Extraterrestrial Physics, New Mexico State University, New York University, Ohio State University, Pennsylvania State University, University of Portsmouth, Princeton University, the Spanish Participation Group, University of Tokyo, University of Utah, Vanderbilt University, University of Virginia, University of Washington, and Yale University.

\bibliography{lit}

\end{document}

%% file: HEADER_prl.tex
\newcommand{\lsim}{\mbox{${\,\hbox{\hbox{$ < $}\kern -0.8em \lower 1.0ex\hbox{$\sim$}}\,}$}}
\newcommand{\gsim}{\mbox{${\,\hbox{\hbox{$ > $}\kern -0.8em \lower 1.0ex\hbox{$\sim$}}\,}$}}

\def\beqn{\vspace{2mm}
\begin{eqnarray}} 
\def\eeqn{\vspace{2mm} 
\end{eqnarray}}

\newcommand{\be}{\begin{equation}}
\newcommand{\ee}{\end{equation}}
\newcommand{\ba}{\begin{eqnarray}}
\newcommand{\ea}{\end{eqnarray}}
\newcommand{\brr}{\begin{array}}
 
\newcommand{\err}{\end{array}}
\newcommand{\bc}{\begin{center}}
\newcommand{\ec}{\end{center}}

\newcommand{\mnras}{MNRAS}
\newcommand{\apjs}{Rev.Astr.Astrphys.}

\newcommand{\jcap}{J.Cosm.Astr.Phys.}
\newcommand{\aap}{Astr.Astrophy.}

\newcommand{\ass}{Astr.Spa.Sci.}
\newcommand{\apjl}{ApJ}

\newcommand{\aj}{AJ}

%% file: voidbao_dr11.bbl
\begin{thebibliography}{120}%
\makeatletter
\providecommand \@ifxundefined [1]{%
 \@ifx{#1\undefined}
}%
\providecommand \@ifnum [1]{%
 \ifnum #1\expandafter \@firstoftwo
 \else \expandafter \@secondoftwo
 \fi
}%
\providecommand \@ifx [1]{%
 \ifx #1\expandafter \@firstoftwo
 \else \expandafter \@secondoftwo
 \fi
}%
\providecommand \natexlab [1]{#1}%
\providecommand \enquote  [1]{``#1''}%
\providecommand \bibnamefont  [1]{#1}%
\providecommand \bibfnamefont [1]{#1}%
\providecommand \citenamefont [1]{#1}%
\providecommand \href@noop [0]{\@secondoftwo}%
\providecommand \href [0]{\begingroup \@sanitize@url \@href}%
\providecommand \@href[1]{\@@startlink{#1}\@@href}%
\providecommand \@@href[1]{\endgroup#1\@@endlink}%
\providecommand \@sanitize@url [0]{\catcode `\\12\catcode `\$12\catcode
  `\&12\catcode `\#12\catcode `\^12\catcode `\_12\catcode `\%12\relax}%
\providecommand \@@startlink[1]{}%
\providecommand \@@endlink[0]{}%
\providecommand \url  [0]{\begingroup\@sanitize@url \@url }%
\providecommand \@url [1]{\endgroup\@href {#1}{\urlprefix }}%
\providecommand \urlprefix  [0]{URL }%
\providecommand \Eprint [0]{\href }%
\providecommand \doibase [0]{http://dx.doi.org/}%
\providecommand \selectlanguage [0]{\@gobble}%
\providecommand \bibinfo  [0]{\@secondoftwo}%
\providecommand \bibfield  [0]{\@secondoftwo}%
\providecommand \translation [1]{[#1]}%
\providecommand \BibitemOpen [0]{}%
\providecommand \bibitemStop [0]{}%
\providecommand \bibitemNoStop [0]{.\EOS\space}%
\providecommand \EOS [0]{\spacefactor3000\relax}%
\providecommand \BibitemShut  [1]{\csname bibitem#1\endcsname}%
\let\auto@bib@innerbib\@empty
%</preamble>
\bibitem [{\citenamefont {{Peebles}}\ and\ \citenamefont {{Yu}}(1970)}]{PY70}%
  \BibitemOpen
  \bibfield  {author} {\bibinfo {author} {\bibfnamefont {P.~J.~E.}\
  \bibnamefont {{Peebles}}}\ and\ \bibinfo {author} {\bibfnamefont {J.~T.}\
  \bibnamefont {{Yu}}},\ }\href {\doibase 10.1086/150713} {\bibfield  {journal}
  {\bibinfo  {journal} {\apj}\ }\textbf {\bibinfo {volume} {162}},\ \bibinfo
  {pages} {815} (\bibinfo {year} {1970})}\BibitemShut {NoStop}%
\bibitem [{\citenamefont {{Sunyaev}}\ and\ \citenamefont
  {{Zeldovich}}(1970)}]{SZ70}%
  \BibitemOpen
  \bibfield  {author} {\bibinfo {author} {\bibfnamefont {R.~A.}\ \bibnamefont
  {{Sunyaev}}}\ and\ \bibinfo {author} {\bibfnamefont {Y.~B.}\ \bibnamefont
  {{Zeldovich}}},\ }\href {\doibase 10.1007/BF00653471} {\bibfield  {journal}
  {\bibinfo  {journal} {\ass}\ }\textbf {\bibinfo {volume} {7}},\ \bibinfo
  {pages} {3} (\bibinfo {year} {1970})}\BibitemShut {NoStop}%
\bibitem [{\citenamefont {{Blake}}\ and\ \citenamefont
  {{Glazebrook}}(2003)}]{BG03}%
  \BibitemOpen
  \bibfield  {author} {\bibinfo {author} {\bibfnamefont {C.}~\bibnamefont
  {{Blake}}}\ and\ \bibinfo {author} {\bibfnamefont {K.}~\bibnamefont
  {{Glazebrook}}},\ }\href {\doibase 10.1086/376983} {\bibfield  {journal}
  {\bibinfo  {journal} {\apj}\ }\textbf {\bibinfo {volume} {594}},\ \bibinfo
  {pages} {665} (\bibinfo {year} {2003})},\ \Eprint
  {http://arxiv.org/abs/astro-ph/0301632} {astro-ph/0301632} \BibitemShut
  {NoStop}%
\bibitem [{\citenamefont {{Seo}}\ and\ \citenamefont
  {{Eisenstein}}(2005)}]{SE05}%
  \BibitemOpen
  \bibfield  {author} {\bibinfo {author} {\bibfnamefont {H.-J.}\ \bibnamefont
  {{Seo}}}\ and\ \bibinfo {author} {\bibfnamefont {D.~J.}\ \bibnamefont
  {{Eisenstein}}},\ }\href {\doibase 10.1086/491599} {\bibfield  {journal}
  {\bibinfo  {journal} {\apj}\ }\textbf {\bibinfo {volume} {633}},\ \bibinfo
  {pages} {575} (\bibinfo {year} {2005})},\ \Eprint
  {http://arxiv.org/abs/astro-ph/0507338} {astro-ph/0507338} \BibitemShut
  {NoStop}%
\bibitem [{\citenamefont {{Spergel et al.}}(2003)}]{WMAP103}%
  \BibitemOpen
  \bibfield  {author} {\bibinfo {author} {\bibfnamefont {D.~N.}\ \bibnamefont
  {{Spergel et al.}}},\ }\href {\doibase 10.1086/377226} {\bibfield  {journal}
  {\bibinfo  {journal} {\apjs}\ }\textbf {\bibinfo {volume} {148}},\ \bibinfo
  {pages} {175} (\bibinfo {year} {2003})},\ \Eprint
  {http://arxiv.org/abs/astro-ph/0302209} {astro-ph/0302209} \BibitemShut
  {NoStop}%
\bibitem [{\citenamefont {{Komatsu et al.}}(2011)}]{WMAP711}%
  \BibitemOpen
  \bibfield  {author} {\bibinfo {author} {\bibfnamefont {E.}~\bibnamefont
  {{Komatsu et al.}}},\ }\href {\doibase 10.1088/0067-0049/192/2/18} {\bibfield
   {journal} {\bibinfo  {journal} {\apjs}\ }\textbf {\bibinfo {volume} {192}},\
  \bibinfo {eid} {18} (\bibinfo {year} {2011})},\ \Eprint
  {http://arxiv.org/abs/1001.4538} {arXiv:1001.4538 [astro-ph.CO]} \BibitemShut
  {NoStop}%
\bibitem [{\citenamefont {{Hinshaw et al.}}(2013)}]{WMAP913}%
  \BibitemOpen
  \bibfield  {author} {\bibinfo {author} {\bibfnamefont {G.}~\bibnamefont
  {{Hinshaw et al.}}},\ }\href {\doibase 10.1088/0067-0049/208/2/19} {\bibfield
   {journal} {\bibinfo  {journal} {\apjs}\ }\textbf {\bibinfo {volume} {208}},\
  \bibinfo {eid} {19} (\bibinfo {year} {2013})},\ \Eprint
  {http://arxiv.org/abs/1212.5226} {arXiv:1212.5226} \BibitemShut {NoStop}%
\bibitem [{\citenamefont {{Planck
  Collaboration}}(2014{\natexlab{a}})}]{PLANCKBAO14}%
  \BibitemOpen
  \bibfield  {author} {\bibinfo {author} {\bibnamefont {{Planck
  Collaboration}}},\ }\href {\doibase 10.1051/0004-6361/201321529} {\bibfield
  {journal} {\bibinfo  {journal} {\aap}\ }\textbf {\bibinfo {volume} {571}},\
  \bibinfo {eid} {A1} (\bibinfo {year} {2014}{\natexlab{a}})},\ \Eprint
  {http://arxiv.org/abs/1303.5062} {arXiv:1303.5062} \BibitemShut {NoStop}%
\bibitem [{\citenamefont {{Cole et al.}}(2005)}]{CPP05}%
  \BibitemOpen
  \bibfield  {author} {\bibinfo {author} {\bibfnamefont {S.}~\bibnamefont
  {{Cole et al.}}},\ }\href {\doibase 10.1111/j.1365-2966.2005.09318.x}
  {\bibfield  {journal} {\bibinfo  {journal} {\mnras}\ }\textbf {\bibinfo
  {volume} {362}},\ \bibinfo {pages} {505} (\bibinfo {year} {2005})},\ \Eprint
  {http://arxiv.org/abs/astro-ph/0501174} {astro-ph/0501174} \BibitemShut
  {NoStop}%
\bibitem [{\citenamefont {{Eisenstein et al.}}(2005)}]{EZH05}%
  \BibitemOpen
  \bibfield  {author} {\bibinfo {author} {\bibfnamefont {D.~J.}\ \bibnamefont
  {{Eisenstein et al.}}},\ }\href {\doibase 10.1086/466512} {\bibfield
  {journal} {\bibinfo  {journal} {\apj}\ }\textbf {\bibinfo {volume} {633}},\
  \bibinfo {pages} {560} (\bibinfo {year} {2005})},\ \Eprint
  {http://arxiv.org/abs/astro-ph/0501171} {astro-ph/0501171} \BibitemShut
  {NoStop}%
\bibitem [{\citenamefont {{Percival et al.}}(2010)}]{PRE10}%
  \BibitemOpen
  \bibfield  {author} {\bibinfo {author} {\bibfnamefont {W.~J.}\ \bibnamefont
  {{Percival et al.}}},\ }\href {\doibase 10.1111/j.1365-2966.2009.15812.x}
  {\bibfield  {journal} {\bibinfo  {journal} {\mnras}\ }\textbf {\bibinfo
  {volume} {401}},\ \bibinfo {pages} {2148} (\bibinfo {year} {2010})},\ \Eprint
  {http://arxiv.org/abs/0907.1660} {arXiv:0907.1660 [astro-ph.CO]} \BibitemShut
  {NoStop}%
\bibitem [{\citenamefont {{Blake et al.}}(2011)}]{BKB11}%
  \BibitemOpen
  \bibfield  {author} {\bibinfo {author} {\bibfnamefont {C.}~\bibnamefont
  {{Blake et al.}}},\ }\href {\doibase 10.1111/j.1365-2966.2011.19592.x}
  {\bibfield  {journal} {\bibinfo  {journal} {\mnras}\ }\textbf {\bibinfo
  {volume} {418}},\ \bibinfo {pages} {1707} (\bibinfo {year} {2011})},\ \Eprint
  {http://arxiv.org/abs/1108.2635} {arXiv:1108.2635} \BibitemShut {NoStop}%
\bibitem [{\citenamefont {{Beutler et al.}}(2011)}]{BBC11}%
  \BibitemOpen
  \bibfield  {author} {\bibinfo {author} {\bibfnamefont {F.}~\bibnamefont
  {{Beutler et al.}}},\ }\href {\doibase 10.1111/j.1365-2966.2011.19250.x}
  {\bibfield  {journal} {\bibinfo  {journal} {\mnras}\ }\textbf {\bibinfo
  {volume} {416}},\ \bibinfo {pages} {3017} (\bibinfo {year} {2011})},\ \Eprint
  {http://arxiv.org/abs/1106.3366} {arXiv:1106.3366} \BibitemShut {NoStop}%
\bibitem [{\citenamefont {{Anderson et al.}}(2014)}]{AAB14}%
  \BibitemOpen
  \bibfield  {author} {\bibinfo {author} {\bibfnamefont {L.}~\bibnamefont
  {{Anderson et al.}}},\ }\href {\doibase 10.1093/mnras/stu523} {\bibfield
  {journal} {\bibinfo  {journal} {\mnras}\ }\textbf {\bibinfo {volume} {441}},\
  \bibinfo {pages} {24} (\bibinfo {year} {2014})},\ \Eprint
  {http://arxiv.org/abs/1312.4877} {arXiv:1312.4877} \BibitemShut {NoStop}%
\bibitem [{\citenamefont {{Busca et al.}}(2013)}]{BDR13}%
  \BibitemOpen
  \bibfield  {author} {\bibinfo {author} {\bibfnamefont {N.~G.}\ \bibnamefont
  {{Busca et al.}}},\ }\href {\doibase 10.1051/0004-6361/201220724} {\bibfield
  {journal} {\bibinfo  {journal} {\aap}\ }\textbf {\bibinfo {volume} {552}},\
  \bibinfo {eid} {A96} (\bibinfo {year} {2013})},\ \Eprint
  {http://arxiv.org/abs/1211.2616} {arXiv:1211.2616 [astro-ph.CO]} \BibitemShut
  {NoStop}%
\bibitem [{\citenamefont {{Slosar et al.}}(2013)}]{SIK13}%
  \BibitemOpen
  \bibfield  {author} {\bibinfo {author} {\bibfnamefont {A.}~\bibnamefont
  {{Slosar et al.}}},\ }\href {\doibase 10.1088/1475-7516/2013/04/026}
  {\bibfield  {journal} {\bibinfo  {journal} {\jcap}\ }\textbf {\bibinfo
  {volume} {4}},\ \bibinfo {eid} {026} (\bibinfo {year} {2013})},\ \Eprint
  {http://arxiv.org/abs/1301.3459} {arXiv:1301.3459} \BibitemShut {NoStop}%
\bibitem [{\citenamefont {{Delubac et al.}}(2015)}]{DBB15}%
  \BibitemOpen
  \bibfield  {author} {\bibinfo {author} {\bibfnamefont {T.}~\bibnamefont
  {{Delubac et al.}}},\ }\href {\doibase 10.1051/0004-6361/201423969}
  {\bibfield  {journal} {\bibinfo  {journal} {\aap}\ }\textbf {\bibinfo
  {volume} {574}},\ \bibinfo {eid} {A59} (\bibinfo {year} {2015})},\ \Eprint
  {http://arxiv.org/abs/1404.1801} {arXiv:1404.1801} \BibitemShut {NoStop}%
\bibitem [{\citenamefont {{Aubourg et al.}}(2014)}]{ABB14}%
  \BibitemOpen
  \bibfield  {author} {\bibinfo {author} {\bibfnamefont {{\'E}.}~\bibnamefont
  {{Aubourg et al.}}},\ }\href@noop {} {\bibfield  {journal} {\bibinfo
  {journal} {ArXiv e-prints}\ } (\bibinfo {year} {2014})},\ \Eprint
  {http://arxiv.org/abs/1411.1074} {arXiv:1411.1074} \BibitemShut {NoStop}%
\bibitem [{\citenamefont {{Colless et al.}}(2001)}]{CDM01}%
  \BibitemOpen
  \bibfield  {author} {\bibinfo {author} {\bibfnamefont {M.}~\bibnamefont
  {{Colless et al.}}},\ }\href {\doibase 10.1046/j.1365-8711.2001.04902.x}
  {\bibfield  {journal} {\bibinfo  {journal} {\mnras}\ }\textbf {\bibinfo
  {volume} {328}},\ \bibinfo {pages} {1039} (\bibinfo {year} {2001})},\ \Eprint
  {http://arxiv.org/abs/astro-ph/0106498} {astro-ph/0106498} \BibitemShut
  {NoStop}%
\bibitem [{\citenamefont {{York}}\ and\ \citenamefont {{SDSS
  Collaboration}}(2000)}]{SDSS00}%
  \BibitemOpen
  \bibfield  {author} {\bibinfo {author} {\bibfnamefont {D.~G.}\ \bibnamefont
  {{York}}}\ and\ \bibinfo {author} {\bibnamefont {{SDSS Collaboration}}},\
  }\href {\doibase 10.1086/301513} {\bibfield  {journal} {\bibinfo  {journal}
  {\aj}\ }\textbf {\bibinfo {volume} {120}},\ \bibinfo {pages} {1579} (\bibinfo
  {year} {2000})},\ \Eprint {http://arxiv.org/abs/astro-ph/0006396}
  {astro-ph/0006396} \BibitemShut {NoStop}%
\bibitem [{\citenamefont {{Drinkwater et al.}}(2010)}]{wigglez2010}%
  \BibitemOpen
  \bibfield  {author} {\bibinfo {author} {\bibfnamefont {M.~J.}\ \bibnamefont
  {{Drinkwater et al.}}},\ }\href {\doibase 10.1111/j.1365-2966.2009.15754.x}
  {\bibfield  {journal} {\bibinfo  {journal} {\mnras}\ }\textbf {\bibinfo
  {volume} {401}},\ \bibinfo {pages} {1429} (\bibinfo {year} {2010})},\ \Eprint
  {http://arxiv.org/abs/0911.4246} {arXiv:0911.4246 [astro-ph.CO]} \BibitemShut
  {NoStop}%
\bibitem [{\citenamefont {{White et al.}}(2011)}]{boss2011}%
  \BibitemOpen
  \bibfield  {author} {\bibinfo {author} {\bibfnamefont {M.}~\bibnamefont
  {{White et al.}}},\ }\href {\doibase 10.1088/0004-637X/728/2/126} {\bibfield
  {journal} {\bibinfo  {journal} {\apj}\ }\textbf {\bibinfo {volume} {728}},\
  \bibinfo {eid} {126} (\bibinfo {year} {2011})},\ \Eprint
  {http://arxiv.org/abs/1010.4915} {arXiv:1010.4915 [astro-ph.CO]} \BibitemShut
  {NoStop}%
\bibitem [{\citenamefont {{Schlegel et al.}}(2011)}]{bigboss2011}%
  \BibitemOpen
  \bibfield  {author} {\bibinfo {author} {\bibfnamefont {D.}~\bibnamefont
  {{Schlegel et al.}}},\ }\href@noop {} {\bibfield  {journal} {\bibinfo
  {journal} {ArXiv e-prints}\ } (\bibinfo {year} {2011})},\ \Eprint
  {http://arxiv.org/abs/1106.1706} {arXiv:1106.1706 [astro-ph.IM]} \BibitemShut
  {NoStop}%
\bibitem [{\citenamefont {{Albrecht}}\ \emph {et~al.}(2006)\citenamefont
  {{Albrecht}}, \citenamefont {{Bernstein}}, \citenamefont {{Cahn}},
  \citenamefont {{Freedman}}, \citenamefont {{Hewitt}}, \citenamefont {{Hu}},
  \citenamefont {{Huth}}, \citenamefont {{Kamionkowski}}, \citenamefont
  {{Kolb}}, \citenamefont {{Knox}}, \citenamefont {{Mather}}, \citenamefont
  {{Staggs}},\ and\ \citenamefont {{Suntzeff}}}]{2006astro.ph..9591A}%
  \BibitemOpen
  \bibfield  {author} {\bibinfo {author} {\bibfnamefont {A.}~\bibnamefont
  {{Albrecht}}}, \bibinfo {author} {\bibfnamefont {G.}~\bibnamefont
  {{Bernstein}}}, \bibinfo {author} {\bibfnamefont {R.}~\bibnamefont {{Cahn}}},
  \bibinfo {author} {\bibfnamefont {W.~L.}\ \bibnamefont {{Freedman}}},
  \bibinfo {author} {\bibfnamefont {J.}~\bibnamefont {{Hewitt}}}, \bibinfo
  {author} {\bibfnamefont {W.}~\bibnamefont {{Hu}}}, \bibinfo {author}
  {\bibfnamefont {J.}~\bibnamefont {{Huth}}}, \bibinfo {author} {\bibfnamefont
  {M.}~\bibnamefont {{Kamionkowski}}}, \bibinfo {author} {\bibfnamefont
  {E.~W.}\ \bibnamefont {{Kolb}}}, \bibinfo {author} {\bibfnamefont
  {L.}~\bibnamefont {{Knox}}}, \bibinfo {author} {\bibfnamefont {J.~C.}\
  \bibnamefont {{Mather}}}, \bibinfo {author} {\bibfnamefont {S.}~\bibnamefont
  {{Staggs}}}, \ and\ \bibinfo {author} {\bibfnamefont {N.~B.}\ \bibnamefont
  {{Suntzeff}}},\ }\href@noop {} {\bibfield  {journal} {\bibinfo  {journal}
  {ArXiv Astrophysics e-prints}\ } (\bibinfo {year} {2006})},\ \Eprint
  {http://arxiv.org/abs/astro-ph/0609591} {astro-ph/0609591} \BibitemShut
  {NoStop}%
\bibitem [{\citenamefont {{LSST Dark Energy Science
  Collaboration}}(2012)}]{lsst2012}%
  \BibitemOpen
  \bibfield  {author} {\bibinfo {author} {\bibnamefont {{LSST Dark Energy
  Science Collaboration}}},\ }\href@noop {} {\bibfield  {journal} {\bibinfo
  {journal} {ArXiv e-prints}\ } (\bibinfo {year} {2012})},\ \Eprint
  {http://arxiv.org/abs/1211.0310} {arXiv:1211.0310 [astro-ph.CO]} \BibitemShut
  {NoStop}%
\bibitem [{\citenamefont {{Benitez et al.}}(2014)}]{jpas2014}%
  \BibitemOpen
  \bibfield  {author} {\bibinfo {author} {\bibfnamefont {N.}~\bibnamefont
  {{Benitez et al.}}},\ }\href@noop {} {\bibfield  {journal} {\bibinfo
  {journal} {ArXiv e-prints}\ } (\bibinfo {year} {2014})},\ \Eprint
  {http://arxiv.org/abs/1403.5237} {arXiv:1403.5237 [astro-ph.CO]} \BibitemShut
  {NoStop}%
\bibitem [{\citenamefont {{de Jong et al.}}(2012)}]{4most}%
  \BibitemOpen
  \bibfield  {author} {\bibinfo {author} {\bibfnamefont {R.~S.}\ \bibnamefont
  {{de Jong et al.}}},\ }in\ \href {\doibase 10.1117/12.926239} {\emph
  {\bibinfo {booktitle} {Society of Photo-Optical Instrumentation Engineers
  (SPIE) Conference Series}}},\ \bibinfo {series} {Society of Photo-Optical
  Instrumentation Engineers (SPIE) Conference Series}, Vol.\ \bibinfo {volume}
  {8446}\ (\bibinfo {year} {2012})\ \Eprint {http://arxiv.org/abs/1206.6885}
  {arXiv:1206.6885 [astro-ph.IM]} \BibitemShut {NoStop}%
\bibitem [{\citenamefont {{Laureijs}}(2009)}]{euclid2009}%
  \BibitemOpen
  \bibfield  {author} {\bibinfo {author} {\bibfnamefont {R.}~\bibnamefont
  {{Laureijs}}},\ }\href@noop {} {\bibfield  {journal} {\bibinfo  {journal}
  {ArXiv e-prints}\ } (\bibinfo {year} {2009})},\ \Eprint
  {http://arxiv.org/abs/0912.0914} {arXiv:0912.0914 [astro-ph.CO]} \BibitemShut
  {NoStop}%
\bibitem [{\citenamefont {{Kirshner}}\ \emph {et~al.}(1981)\citenamefont
  {{Kirshner}}, \citenamefont {{Oemler}}, \citenamefont {{Schechter}},\ and\
  \citenamefont {{Shectman}}}]{KOS81}%
  \BibitemOpen
  \bibfield  {author} {\bibinfo {author} {\bibfnamefont {R.~P.}\ \bibnamefont
  {{Kirshner}}}, \bibinfo {author} {\bibfnamefont {A.}~\bibnamefont {{Oemler}},
  \bibfnamefont {Jr.}}, \bibinfo {author} {\bibfnamefont {P.~L.}\ \bibnamefont
  {{Schechter}}}, \ and\ \bibinfo {author} {\bibfnamefont {S.~A.}\ \bibnamefont
  {{Shectman}}},\ }\href {\doibase 10.1086/183623} {\bibfield  {journal}
  {\bibinfo  {journal} {\apjl}\ }\textbf {\bibinfo {volume} {248}},\ \bibinfo
  {pages} {L57} (\bibinfo {year} {1981})}\BibitemShut {NoStop}%
\bibitem [{\citenamefont {{Klypin}}\ and\ \citenamefont
  {{Shandarin}}(1983)}]{Klypin1983}%
  \BibitemOpen
  \bibfield  {author} {\bibinfo {author} {\bibfnamefont {A.~A.}\ \bibnamefont
  {{Klypin}}}\ and\ \bibinfo {author} {\bibfnamefont {S.~F.}\ \bibnamefont
  {{Shandarin}}},\ }\href@noop {} {\bibfield  {journal} {\bibinfo  {journal}
  {\mnras}\ }\textbf {\bibinfo {volume} {204}},\ \bibinfo {pages} {891}
  (\bibinfo {year} {1983})}\BibitemShut {NoStop}%
\bibitem [{\citenamefont {{Blumenthal}}\ \emph {et~al.}(1984)\citenamefont
  {{Blumenthal}}, \citenamefont {{Faber}}, \citenamefont {{Primack}},\ and\
  \citenamefont {{Rees}}}]{Blumenthal1984}%
  \BibitemOpen
  \bibfield  {author} {\bibinfo {author} {\bibfnamefont {G.~R.}\ \bibnamefont
  {{Blumenthal}}}, \bibinfo {author} {\bibfnamefont {S.~M.}\ \bibnamefont
  {{Faber}}}, \bibinfo {author} {\bibfnamefont {J.~R.}\ \bibnamefont
  {{Primack}}}, \ and\ \bibinfo {author} {\bibfnamefont {M.~J.}\ \bibnamefont
  {{Rees}}},\ }\href {\doibase 10.1038/311517a0} {\bibfield  {journal}
  {\bibinfo  {journal} {\nat}\ }\textbf {\bibinfo {volume} {311}},\ \bibinfo
  {pages} {517} (\bibinfo {year} {1984})}\BibitemShut {NoStop}%
\bibitem [{\citenamefont {{Davis}}\ \emph {et~al.}(1985)\citenamefont
  {{Davis}}, \citenamefont {{Efstathiou}}, \citenamefont {{Frenk}},\ and\
  \citenamefont {{White}}}]{DEF85}%
  \BibitemOpen
  \bibfield  {author} {\bibinfo {author} {\bibfnamefont {M.}~\bibnamefont
  {{Davis}}}, \bibinfo {author} {\bibfnamefont {G.}~\bibnamefont
  {{Efstathiou}}}, \bibinfo {author} {\bibfnamefont {C.~S.}\ \bibnamefont
  {{Frenk}}}, \ and\ \bibinfo {author} {\bibfnamefont {S.~D.~M.}\ \bibnamefont
  {{White}}},\ }\href {\doibase 10.1086/163168} {\bibfield  {journal} {\bibinfo
   {journal} {\apj}\ }\textbf {\bibinfo {volume} {292}},\ \bibinfo {pages}
  {371} (\bibinfo {year} {1985})}\BibitemShut {NoStop}%
\bibitem [{\citenamefont {{de Lapparent}}\ \emph {et~al.}(1986)\citenamefont
  {{de Lapparent}}, \citenamefont {{Geller}},\ and\ \citenamefont
  {{Huchra}}}]{LGH86}%
  \BibitemOpen
  \bibfield  {author} {\bibinfo {author} {\bibfnamefont {V.}~\bibnamefont {{de
  Lapparent}}}, \bibinfo {author} {\bibfnamefont {M.~J.}\ \bibnamefont
  {{Geller}}}, \ and\ \bibinfo {author} {\bibfnamefont {J.~P.}\ \bibnamefont
  {{Huchra}}},\ }\href {\doibase 10.1086/184625} {\bibfield  {journal}
  {\bibinfo  {journal} {\apjl}\ }\textbf {\bibinfo {volume} {302}},\ \bibinfo
  {pages} {L1} (\bibinfo {year} {1986})}\BibitemShut {NoStop}%
\bibitem [{\citenamefont {{Vogeley}}\ \emph {et~al.}(1994)\citenamefont
  {{Vogeley}}, \citenamefont {{Geller}}, \citenamefont {{Park}},\ and\
  \citenamefont {{Huchra}}}]{VGP94}%
  \BibitemOpen
  \bibfield  {author} {\bibinfo {author} {\bibfnamefont {M.~S.}\ \bibnamefont
  {{Vogeley}}}, \bibinfo {author} {\bibfnamefont {M.~J.}\ \bibnamefont
  {{Geller}}}, \bibinfo {author} {\bibfnamefont {C.}~\bibnamefont {{Park}}}, \
  and\ \bibinfo {author} {\bibfnamefont {J.~P.}\ \bibnamefont {{Huchra}}},\
  }\href {\doibase 10.1086/117110} {\bibfield  {journal} {\bibinfo  {journal}
  {\aj}\ }\textbf {\bibinfo {volume} {108}},\ \bibinfo {pages} {745} (\bibinfo
  {year} {1994})}\BibitemShut {NoStop}%
\bibitem [{\citenamefont {{El-Ad}}\ and\ \citenamefont {{Piran}}(1997)}]{EP97}%
  \BibitemOpen
  \bibfield  {author} {\bibinfo {author} {\bibfnamefont {H.}~\bibnamefont
  {{El-Ad}}}\ and\ \bibinfo {author} {\bibfnamefont {T.}~\bibnamefont
  {{Piran}}},\ }\href@noop {} {\bibfield  {journal} {\bibinfo  {journal}
  {\apj}\ }\textbf {\bibinfo {volume} {491}},\ \bibinfo {pages} {421} (\bibinfo
  {year} {1997})},\ \Eprint {http://arxiv.org/abs/astro-ph/9702135}
  {astro-ph/9702135} \BibitemShut {NoStop}%
\bibitem [{\citenamefont {{M{\"u}ller}}\ \emph {et~al.}(2000)\citenamefont
  {{M{\"u}ller}}, \citenamefont {{Arbabi-Bidgoli}}, \citenamefont {{Einasto}},\
  and\ \citenamefont {{Tucker}}}]{MAE00}%
  \BibitemOpen
  \bibfield  {author} {\bibinfo {author} {\bibfnamefont {V.}~\bibnamefont
  {{M{\"u}ller}}}, \bibinfo {author} {\bibfnamefont {S.}~\bibnamefont
  {{Arbabi-Bidgoli}}}, \bibinfo {author} {\bibfnamefont {J.}~\bibnamefont
  {{Einasto}}}, \ and\ \bibinfo {author} {\bibfnamefont {D.}~\bibnamefont
  {{Tucker}}},\ }\href {\doibase 10.1046/j.1365-8711.2000.03775.x} {\bibfield
  {journal} {\bibinfo  {journal} {\mnras}\ }\textbf {\bibinfo {volume} {318}},\
  \bibinfo {pages} {280} (\bibinfo {year} {2000})},\ \Eprint
  {http://arxiv.org/abs/astro-ph/0005063} {astro-ph/0005063} \BibitemShut
  {NoStop}%
\bibitem [{\citenamefont {{Plionis}}\ and\ \citenamefont
  {{Basilakos}}(2002)}]{PB02}%
  \BibitemOpen
  \bibfield  {author} {\bibinfo {author} {\bibfnamefont {M.}~\bibnamefont
  {{Plionis}}}\ and\ \bibinfo {author} {\bibfnamefont {S.}~\bibnamefont
  {{Basilakos}}},\ }\href {\doibase 10.1046/j.1365-8711.2002.05069.x}
  {\bibfield  {journal} {\bibinfo  {journal} {\mnras}\ }\textbf {\bibinfo
  {volume} {330}},\ \bibinfo {pages} {399} (\bibinfo {year} {2002})},\ \Eprint
  {http://arxiv.org/abs/astro-ph/0106491} {astro-ph/0106491} \BibitemShut
  {NoStop}%
\bibitem [{\citenamefont {{Croton et al.}}(2004)}]{CCG04}%
  \BibitemOpen
  \bibfield  {author} {\bibinfo {author} {\bibfnamefont {D.~J.}\ \bibnamefont
  {{Croton et al.}}},\ }\href {\doibase 10.1111/j.1365-2966.2004.07968.x}
  {\bibfield  {journal} {\bibinfo  {journal} {\mnras}\ }\textbf {\bibinfo
  {volume} {352}},\ \bibinfo {pages} {828} (\bibinfo {year} {2004})},\ \Eprint
  {http://arxiv.org/abs/astro-ph/0401406} {astro-ph/0401406} \BibitemShut
  {NoStop}%
\bibitem [{\citenamefont {{Hoyle}}\ and\ \citenamefont
  {{Vogeley}}(2004)}]{HV04}%
  \BibitemOpen
  \bibfield  {author} {\bibinfo {author} {\bibfnamefont {F.}~\bibnamefont
  {{Hoyle}}}\ and\ \bibinfo {author} {\bibfnamefont {M.~S.}\ \bibnamefont
  {{Vogeley}}},\ }\href {\doibase 10.1086/386279} {\bibfield  {journal}
  {\bibinfo  {journal} {\apj}\ }\textbf {\bibinfo {volume} {607}},\ \bibinfo
  {pages} {751} (\bibinfo {year} {2004})},\ \Eprint
  {http://arxiv.org/abs/astro-ph/0312533} {astro-ph/0312533} \BibitemShut
  {NoStop}%
\bibitem [{\citenamefont {{Patiri}}\ \emph
  {et~al.}(2006{\natexlab{a}})\citenamefont {{Patiri}}, \citenamefont
  {{Betancort-Rijo}}, \citenamefont {{Prada}}, \citenamefont {{Klypin}},\ and\
  \citenamefont {{Gottl{\"o}ber}}}]{PBP06}%
  \BibitemOpen
  \bibfield  {author} {\bibinfo {author} {\bibfnamefont {S.~G.}\ \bibnamefont
  {{Patiri}}}, \bibinfo {author} {\bibfnamefont {J.~E.}\ \bibnamefont
  {{Betancort-Rijo}}}, \bibinfo {author} {\bibfnamefont {F.}~\bibnamefont
  {{Prada}}}, \bibinfo {author} {\bibfnamefont {A.}~\bibnamefont {{Klypin}}}, \
  and\ \bibinfo {author} {\bibfnamefont {S.}~\bibnamefont {{Gottl{\"o}ber}}},\
  }\href {\doibase 10.1111/j.1365-2966.2006.10305.x} {\bibfield  {journal}
  {\bibinfo  {journal} {\mnras}\ }\textbf {\bibinfo {volume} {369}},\ \bibinfo
  {pages} {335} (\bibinfo {year} {2006}{\natexlab{a}})},\ \Eprint
  {http://arxiv.org/abs/astro-ph/0506668} {astro-ph/0506668} \BibitemShut
  {NoStop}%
\bibitem [{\citenamefont {{Conroy et al.}}(2005)}]{CCW05}%
  \BibitemOpen
  \bibfield  {author} {\bibinfo {author} {\bibfnamefont {C.}~\bibnamefont
  {{Conroy et al.}}},\ }\href {\doibase 10.1086/497682} {\bibfield  {journal}
  {\bibinfo  {journal} {\apj}\ }\textbf {\bibinfo {volume} {635}},\ \bibinfo
  {pages} {990} (\bibinfo {year} {2005})},\ \Eprint
  {http://arxiv.org/abs/astro-ph/0508250} {astro-ph/0508250} \BibitemShut
  {NoStop}%
\bibitem [{\citenamefont {{Nuza}}\ \emph {et~al.}(2014)\citenamefont {{Nuza}},
  \citenamefont {{Kitaura}}, \citenamefont {{He{\ss}}}, \citenamefont
  {{Libeskind}},\ and\ \citenamefont {{M{\"u}ller}}}]{NKH14}%
  \BibitemOpen
  \bibfield  {author} {\bibinfo {author} {\bibfnamefont {S.~E.}\ \bibnamefont
  {{Nuza}}}, \bibinfo {author} {\bibfnamefont {F.-S.}\ \bibnamefont
  {{Kitaura}}}, \bibinfo {author} {\bibfnamefont {S.}~\bibnamefont
  {{He{\ss}}}}, \bibinfo {author} {\bibfnamefont {N.~I.}\ \bibnamefont
  {{Libeskind}}}, \ and\ \bibinfo {author} {\bibfnamefont {V.}~\bibnamefont
  {{M{\"u}ller}}},\ }\href {\doibase 10.1093/mnras/stu1746} {\bibfield
  {journal} {\bibinfo  {journal} {\mnras}\ }\textbf {\bibinfo {volume} {445}},\
  \bibinfo {pages} {988} (\bibinfo {year} {2014})},\ \Eprint
  {http://arxiv.org/abs/1406.1004} {arXiv:1406.1004} \BibitemShut {NoStop}%
\bibitem [{\citenamefont {{Platen}}\ \emph {et~al.}(2011)\citenamefont
  {{Platen}}, \citenamefont {{van de Weygaert}}, \citenamefont {{Jones}},
  \citenamefont {{Vegter}},\ and\ \citenamefont {{Calvo}}}]{PWJ11}%
  \BibitemOpen
  \bibfield  {author} {\bibinfo {author} {\bibfnamefont {E.}~\bibnamefont
  {{Platen}}}, \bibinfo {author} {\bibfnamefont {R.}~\bibnamefont {{van de
  Weygaert}}}, \bibinfo {author} {\bibfnamefont {B.~J.~T.}\ \bibnamefont
  {{Jones}}}, \bibinfo {author} {\bibfnamefont {G.}~\bibnamefont {{Vegter}}}, \
  and\ \bibinfo {author} {\bibfnamefont {M.~A.~A.}\ \bibnamefont {{Calvo}}},\
  }\href {\doibase 10.1111/j.1365-2966.2011.18905.x} {\bibfield  {journal}
  {\bibinfo  {journal} {\mnras}\ }\textbf {\bibinfo {volume} {416}},\ \bibinfo
  {pages} {2494} (\bibinfo {year} {2011})},\ \Eprint
  {http://arxiv.org/abs/1107.1488} {arXiv:1107.1488 [astro-ph.CO]} \BibitemShut
  {NoStop}%
\bibitem [{\citenamefont {{Varela}}\ \emph {et~al.}(2012)\citenamefont
  {{Varela}}, \citenamefont {{Betancort-Rijo}}, \citenamefont {{Trujillo}},\
  and\ \citenamefont {{Ricciardelli}}}]{VBT12}%
  \BibitemOpen
  \bibfield  {author} {\bibinfo {author} {\bibfnamefont {J.}~\bibnamefont
  {{Varela}}}, \bibinfo {author} {\bibfnamefont {J.}~\bibnamefont
  {{Betancort-Rijo}}}, \bibinfo {author} {\bibfnamefont {I.}~\bibnamefont
  {{Trujillo}}}, \ and\ \bibinfo {author} {\bibfnamefont {E.}~\bibnamefont
  {{Ricciardelli}}},\ }\href {\doibase 10.1088/0004-637X/744/2/82} {\bibfield
  {journal} {\bibinfo  {journal} {\apj}\ }\textbf {\bibinfo {volume} {744}},\
  \bibinfo {eid} {82} (\bibinfo {year} {2012})},\ \Eprint
  {http://arxiv.org/abs/1109.2056} {arXiv:1109.2056} \BibitemShut {NoStop}%
\bibitem [{\citenamefont {{Pan}}\ \emph {et~al.}(2012)\citenamefont {{Pan}},
  \citenamefont {{Vogeley}}, \citenamefont {{Hoyle}}, \citenamefont {{Choi}},\
  and\ \citenamefont {{Park}}}]{PVH12}%
  \BibitemOpen
  \bibfield  {author} {\bibinfo {author} {\bibfnamefont {D.~C.}\ \bibnamefont
  {{Pan}}}, \bibinfo {author} {\bibfnamefont {M.~S.}\ \bibnamefont
  {{Vogeley}}}, \bibinfo {author} {\bibfnamefont {F.}~\bibnamefont {{Hoyle}}},
  \bibinfo {author} {\bibfnamefont {Y.-Y.}\ \bibnamefont {{Choi}}}, \ and\
  \bibinfo {author} {\bibfnamefont {C.}~\bibnamefont {{Park}}},\ }\href
  {\doibase 10.1111/j.1365-2966.2011.20197.x} {\bibfield  {journal} {\bibinfo
  {journal} {\mnras}\ }\textbf {\bibinfo {volume} {421}},\ \bibinfo {pages}
  {926} (\bibinfo {year} {2012})},\ \Eprint {http://arxiv.org/abs/1103.4156}
  {arXiv:1103.4156} \BibitemShut {NoStop}%
\bibitem [{\citenamefont {{Nadathur}}\ and\ \citenamefont
  {{Hotchkiss}}(2014)}]{NH14}%
  \BibitemOpen
  \bibfield  {author} {\bibinfo {author} {\bibfnamefont {S.}~\bibnamefont
  {{Nadathur}}}\ and\ \bibinfo {author} {\bibfnamefont {S.}~\bibnamefont
  {{Hotchkiss}}},\ }\href {\doibase 10.1093/mnras/stu349} {\bibfield  {journal}
  {\bibinfo  {journal} {\mnras}\ }\textbf {\bibinfo {volume} {440}},\ \bibinfo
  {pages} {1248} (\bibinfo {year} {2014})},\ \Eprint
  {http://arxiv.org/abs/1310.2791} {arXiv:1310.2791} \BibitemShut {NoStop}%
\bibitem [{\citenamefont {{Sutter}}\ \emph
  {et~al.}(2014{\natexlab{a}})\citenamefont {{Sutter}}, \citenamefont
  {{Lavaux}}, \citenamefont {{Wandelt}}, \citenamefont {{Weinberg}},
  \citenamefont {{Warren}},\ and\ \citenamefont {{Pisani}}}]{SLW14}%
  \BibitemOpen
  \bibfield  {author} {\bibinfo {author} {\bibfnamefont {P.~M.}\ \bibnamefont
  {{Sutter}}}, \bibinfo {author} {\bibfnamefont {G.}~\bibnamefont {{Lavaux}}},
  \bibinfo {author} {\bibfnamefont {B.~D.}\ \bibnamefont {{Wandelt}}}, \bibinfo
  {author} {\bibfnamefont {D.~H.}\ \bibnamefont {{Weinberg}}}, \bibinfo
  {author} {\bibfnamefont {M.~S.}\ \bibnamefont {{Warren}}}, \ and\ \bibinfo
  {author} {\bibfnamefont {A.}~\bibnamefont {{Pisani}}},\ }\href {\doibase
  10.1093/mnras/stu1094} {\bibfield  {journal} {\bibinfo  {journal} {\mnras}\
  }\textbf {\bibinfo {volume} {442}},\ \bibinfo {pages} {3127} (\bibinfo {year}
  {2014}{\natexlab{a}})},\ \Eprint {http://arxiv.org/abs/1310.7155}
  {arXiv:1310.7155} \BibitemShut {NoStop}%
\bibitem [{\citenamefont {{Beygu}}\ \emph {et~al.}(2015)\citenamefont
  {{Beygu}}, \citenamefont {{Kreckel}}, \citenamefont {{van der Hulst}},
  \citenamefont {{Peletier}}, \citenamefont {{Jarrett}}, \citenamefont {{van de
  Weygaert}}, \citenamefont {{van Gorkom}},\ and\ \citenamefont
  {{Arag{\'o}n-Calvo}}}]{BKH15}%
  \BibitemOpen
  \bibfield  {author} {\bibinfo {author} {\bibfnamefont {B.}~\bibnamefont
  {{Beygu}}}, \bibinfo {author} {\bibfnamefont {K.}~\bibnamefont {{Kreckel}}},
  \bibinfo {author} {\bibfnamefont {J.~M.}\ \bibnamefont {{van der Hulst}}},
  \bibinfo {author} {\bibfnamefont {R.}~\bibnamefont {{Peletier}}}, \bibinfo
  {author} {\bibfnamefont {T.}~\bibnamefont {{Jarrett}}}, \bibinfo {author}
  {\bibfnamefont {R.}~\bibnamefont {{van de Weygaert}}}, \bibinfo {author}
  {\bibfnamefont {J.~H.}\ \bibnamefont {{van Gorkom}}}, \ and\ \bibinfo
  {author} {\bibfnamefont {M.}~\bibnamefont {{Arag{\'o}n-Calvo}}},\ }\href@noop
  {} {\bibfield  {journal} {\bibinfo  {journal} {ArXiv e-prints}\ } (\bibinfo
  {year} {2015})},\ \Eprint {http://arxiv.org/abs/1501.02577}
  {arXiv:1501.02577} \BibitemShut {NoStop}%
\bibitem [{\citenamefont {{Micheletti}}\ \emph {et~al.}(2014)\citenamefont
  {{Micheletti}}, \citenamefont {{Iovino}}, \citenamefont {{Hawken}},
  \citenamefont {{Granett}}, \citenamefont {{Bolzonella}}, \citenamefont
  {{Cappi}}, \citenamefont {{Guzzo}}, \citenamefont {{Abbas}},\ and\
  \citenamefont {et~al.}}]{2014A&A...570A.106M}%
  \BibitemOpen
  \bibfield  {author} {\bibinfo {author} {\bibfnamefont {D.}~\bibnamefont
  {{Micheletti}}}, \bibinfo {author} {\bibfnamefont {A.}~\bibnamefont
  {{Iovino}}}, \bibinfo {author} {\bibfnamefont {A.~J.}\ \bibnamefont
  {{Hawken}}}, \bibinfo {author} {\bibfnamefont {B.~R.}\ \bibnamefont
  {{Granett}}}, \bibinfo {author} {\bibfnamefont {M.}~\bibnamefont
  {{Bolzonella}}}, \bibinfo {author} {\bibfnamefont {A.}~\bibnamefont
  {{Cappi}}}, \bibinfo {author} {\bibfnamefont {L.}~\bibnamefont {{Guzzo}}},
  \bibinfo {author} {\bibfnamefont {U.}~\bibnamefont {{Abbas}}}, \ and\
  \bibinfo {author} {\bibnamefont {et~al.}},\ }\href {\doibase
  10.1051/0004-6361/201424107} {\bibfield  {journal} {\bibinfo  {journal}
  {\aap}\ }\textbf {\bibinfo {volume} {570}},\ \bibinfo {eid} {A106} (\bibinfo
  {year} {2014})},\ \Eprint {http://arxiv.org/abs/1407.2969} {arXiv:1407.2969}
  \BibitemShut {NoStop}%
\bibitem [{\citenamefont {{Betancort-Rijo}}\ \emph {et~al.}(2009)\citenamefont
  {{Betancort-Rijo}}, \citenamefont {{Patiri}}, \citenamefont {{Prada}},\ and\
  \citenamefont {{Romano}}}]{BPP09}%
  \BibitemOpen
  \bibfield  {author} {\bibinfo {author} {\bibfnamefont {J.}~\bibnamefont
  {{Betancort-Rijo}}}, \bibinfo {author} {\bibfnamefont {S.~G.}\ \bibnamefont
  {{Patiri}}}, \bibinfo {author} {\bibfnamefont {F.}~\bibnamefont {{Prada}}}, \
  and\ \bibinfo {author} {\bibfnamefont {A.~E.}\ \bibnamefont {{Romano}}},\
  }\href {\doibase 10.1111/j.1365-2966.2009.15567.x} {\bibfield  {journal}
  {\bibinfo  {journal} {\mnras}\ }\textbf {\bibinfo {volume} {400}},\ \bibinfo
  {pages} {1835} (\bibinfo {year} {2009})},\ \Eprint
  {http://arxiv.org/abs/0901.1609} {arXiv:0901.1609} \BibitemShut {NoStop}%
\bibitem [{\citenamefont {{White}}(1979)}]{W79}%
  \BibitemOpen
  \bibfield  {author} {\bibinfo {author} {\bibfnamefont {S.~D.~M.}\
  \bibnamefont {{White}}},\ }\href@noop {} {\bibfield  {journal} {\bibinfo
  {journal} {\mnras}\ }\textbf {\bibinfo {volume} {186}},\ \bibinfo {pages}
  {145} (\bibinfo {year} {1979})}\BibitemShut {NoStop}%
\bibitem [{\citenamefont {{Politzer}}\ and\ \citenamefont
  {{Preskill}}(1986)}]{PP86}%
  \BibitemOpen
  \bibfield  {author} {\bibinfo {author} {\bibfnamefont {H.~D.}\ \bibnamefont
  {{Politzer}}}\ and\ \bibinfo {author} {\bibfnamefont {J.~P.}\ \bibnamefont
  {{Preskill}}},\ }\href {\doibase 10.1103/PhysRevLett.56.99} {\bibfield
  {journal} {\bibinfo  {journal} {Physical Review Letters}\ }\textbf {\bibinfo
  {volume} {56}},\ \bibinfo {pages} {99} (\bibinfo {year} {1986})}\BibitemShut
  {NoStop}%
\bibitem [{\citenamefont {{Betancort-Rijo}}(1990)}]{B90}%
  \BibitemOpen
  \bibfield  {author} {\bibinfo {author} {\bibfnamefont {J.}~\bibnamefont
  {{Betancort-Rijo}}},\ }\href@noop {} {\bibfield  {journal} {\bibinfo
  {journal} {\mnras}\ }\textbf {\bibinfo {volume} {246}},\ \bibinfo {pages}
  {608} (\bibinfo {year} {1990})}\BibitemShut {NoStop}%
\bibitem [{\citenamefont {{Einasto}}\ \emph {et~al.}(1991)\citenamefont
  {{Einasto}}, \citenamefont {{Einasto}}, \citenamefont {{Gramann}},\ and\
  \citenamefont {{Saar}}}]{EEG91}%
  \BibitemOpen
  \bibfield  {author} {\bibinfo {author} {\bibfnamefont {J.}~\bibnamefont
  {{Einasto}}}, \bibinfo {author} {\bibfnamefont {M.}~\bibnamefont
  {{Einasto}}}, \bibinfo {author} {\bibfnamefont {M.}~\bibnamefont
  {{Gramann}}}, \ and\ \bibinfo {author} {\bibfnamefont {E.}~\bibnamefont
  {{Saar}}},\ }\href@noop {} {\bibfield  {journal} {\bibinfo  {journal}
  {\mnras}\ }\textbf {\bibinfo {volume} {248}},\ \bibinfo {pages} {593}
  (\bibinfo {year} {1991})}\BibitemShut {NoStop}%
\bibitem [{\citenamefont {{Betancort-Rijo}}\ and\ \citenamefont
  {{L{\'o}pez-Corredoira}}(2002)}]{BL02}%
  \BibitemOpen
  \bibfield  {author} {\bibinfo {author} {\bibfnamefont {J.}~\bibnamefont
  {{Betancort-Rijo}}}\ and\ \bibinfo {author} {\bibfnamefont {M.}~\bibnamefont
  {{L{\'o}pez-Corredoira}}},\ }\href {\doibase 10.1086/338328} {\bibfield
  {journal} {\bibinfo  {journal} {\apj}\ }\textbf {\bibinfo {volume} {566}},\
  \bibinfo {pages} {623} (\bibinfo {year} {2002})},\ \Eprint
  {http://arxiv.org/abs/astro-ph/0110624} {astro-ph/0110624} \BibitemShut
  {NoStop}%
\bibitem [{\citenamefont {{Sheth}}\ and\ \citenamefont {{van de
  Weygaert}}(2004)}]{SW04}%
  \BibitemOpen
  \bibfield  {author} {\bibinfo {author} {\bibfnamefont {R.~K.}\ \bibnamefont
  {{Sheth}}}\ and\ \bibinfo {author} {\bibfnamefont {R.}~\bibnamefont {{van de
  Weygaert}}},\ }\href {\doibase 10.1111/j.1365-2966.2004.07661.x} {\bibfield
  {journal} {\bibinfo  {journal} {\mnras}\ }\textbf {\bibinfo {volume} {350}},\
  \bibinfo {pages} {517} (\bibinfo {year} {2004})},\ \Eprint
  {http://arxiv.org/abs/astro-ph/0311260} {astro-ph/0311260} \BibitemShut
  {NoStop}%
\bibitem [{\citenamefont {{Furlanetto}}\ and\ \citenamefont
  {{Piran}}(2006)}]{FP06}%
  \BibitemOpen
  \bibfield  {author} {\bibinfo {author} {\bibfnamefont {S.~R.}\ \bibnamefont
  {{Furlanetto}}}\ and\ \bibinfo {author} {\bibfnamefont {T.}~\bibnamefont
  {{Piran}}},\ }\href {\doibase 10.1111/j.1365-2966.2005.09862.x} {\bibfield
  {journal} {\bibinfo  {journal} {\mnras}\ }\textbf {\bibinfo {volume} {366}},\
  \bibinfo {pages} {467} (\bibinfo {year} {2006})},\ \Eprint
  {http://arxiv.org/abs/astro-ph/0509148} {astro-ph/0509148} \BibitemShut
  {NoStop}%
\bibitem [{\citenamefont {{Paranjape}}\ \emph {et~al.}(2012)\citenamefont
  {{Paranjape}}, \citenamefont {{Lam}},\ and\ \citenamefont {{Sheth}}}]{PLS12}%
  \BibitemOpen
  \bibfield  {author} {\bibinfo {author} {\bibfnamefont {A.}~\bibnamefont
  {{Paranjape}}}, \bibinfo {author} {\bibfnamefont {T.~Y.}\ \bibnamefont
  {{Lam}}}, \ and\ \bibinfo {author} {\bibfnamefont {R.~K.}\ \bibnamefont
  {{Sheth}}},\ }\href {\doibase 10.1111/j.1365-2966.2011.20154.x} {\bibfield
  {journal} {\bibinfo  {journal} {\mnras}\ }\textbf {\bibinfo {volume} {420}},\
  \bibinfo {pages} {1648} (\bibinfo {year} {2012})},\ \Eprint
  {http://arxiv.org/abs/1106.2041} {arXiv:1106.2041 [astro-ph.CO]} \BibitemShut
  {NoStop}%
\bibitem [{\citenamefont {{Jennings}}\ \emph {et~al.}(2013)\citenamefont
  {{Jennings}}, \citenamefont {{Li}},\ and\ \citenamefont {{Hu}}}]{JLH13}%
  \BibitemOpen
  \bibfield  {author} {\bibinfo {author} {\bibfnamefont {E.}~\bibnamefont
  {{Jennings}}}, \bibinfo {author} {\bibfnamefont {Y.}~\bibnamefont {{Li}}}, \
  and\ \bibinfo {author} {\bibfnamefont {W.}~\bibnamefont {{Hu}}},\ }\href
  {\doibase 10.1093/mnras/stt1169} {\bibfield  {journal} {\bibinfo  {journal}
  {\mnras}\ }\textbf {\bibinfo {volume} {434}},\ \bibinfo {pages} {2167}
  (\bibinfo {year} {2013})},\ \Eprint {http://arxiv.org/abs/1304.6087}
  {arXiv:1304.6087} \BibitemShut {NoStop}%
\bibitem [{\citenamefont {{Aragon-Calvo}}\ \emph {et~al.}(2010)\citenamefont
  {{Aragon-Calvo}}, \citenamefont {{van de Weygaert}}, \citenamefont
  {{Araya-Melo}}, \citenamefont {{Platen}},\ and\ \citenamefont
  {{Szalay}}}]{AWA10}%
  \BibitemOpen
  \bibfield  {author} {\bibinfo {author} {\bibfnamefont {M.~A.}\ \bibnamefont
  {{Aragon-Calvo}}}, \bibinfo {author} {\bibfnamefont {R.}~\bibnamefont {{van
  de Weygaert}}}, \bibinfo {author} {\bibfnamefont {P.~A.}\ \bibnamefont
  {{Araya-Melo}}}, \bibinfo {author} {\bibfnamefont {E.}~\bibnamefont
  {{Platen}}}, \ and\ \bibinfo {author} {\bibfnamefont {A.~S.}\ \bibnamefont
  {{Szalay}}},\ }\href {\doibase 10.1111/j.1745-3933.2010.00841.x} {\bibfield
  {journal} {\bibinfo  {journal} {\mnras}\ }\textbf {\bibinfo {volume} {404}},\
  \bibinfo {pages} {L89} (\bibinfo {year} {2010})},\ \Eprint
  {http://arxiv.org/abs/1002.1503} {arXiv:1002.1503} \BibitemShut {NoStop}%
\bibitem [{\citenamefont {{Mathis}}\ and\ \citenamefont
  {{White}}(2002)}]{MW02}%
  \BibitemOpen
  \bibfield  {author} {\bibinfo {author} {\bibfnamefont {H.}~\bibnamefont
  {{Mathis}}}\ and\ \bibinfo {author} {\bibfnamefont {S.~D.~M.}\ \bibnamefont
  {{White}}},\ }\href {\doibase 10.1046/j.1365-8711.2002.06010.x} {\bibfield
  {journal} {\bibinfo  {journal} {\mnras}\ }\textbf {\bibinfo {volume} {337}},\
  \bibinfo {pages} {1193} (\bibinfo {year} {2002})},\ \Eprint
  {http://arxiv.org/abs/astro-ph/0201193} {astro-ph/0201193} \BibitemShut
  {NoStop}%
\bibitem [{\citenamefont {{Benson}}\ \emph {et~al.}(2003)\citenamefont
  {{Benson}}, \citenamefont {{Hoyle}}, \citenamefont {{Torres}},\ and\
  \citenamefont {{Vogeley}}}]{BHT03}%
  \BibitemOpen
  \bibfield  {author} {\bibinfo {author} {\bibfnamefont {A.~J.}\ \bibnamefont
  {{Benson}}}, \bibinfo {author} {\bibfnamefont {F.}~\bibnamefont {{Hoyle}}},
  \bibinfo {author} {\bibfnamefont {F.}~\bibnamefont {{Torres}}}, \ and\
  \bibinfo {author} {\bibfnamefont {M.~S.}\ \bibnamefont {{Vogeley}}},\ }\href
  {\doibase 10.1046/j.1365-8711.2003.06281.x} {\bibfield  {journal} {\bibinfo
  {journal} {\mnras}\ }\textbf {\bibinfo {volume} {340}},\ \bibinfo {pages}
  {160} (\bibinfo {year} {2003})},\ \Eprint
  {http://arxiv.org/abs/astro-ph/0208257} {astro-ph/0208257} \BibitemShut
  {NoStop}%
\bibitem [{\citenamefont {{Blumenthal}}\ \emph {et~al.}(1992)\citenamefont
  {{Blumenthal}}, \citenamefont {{da Costa}}, \citenamefont {{Goldwirth}},
  \citenamefont {{Lecar}},\ and\ \citenamefont {{Piran}}}]{BCG92}%
  \BibitemOpen
  \bibfield  {author} {\bibinfo {author} {\bibfnamefont {G.~R.}\ \bibnamefont
  {{Blumenthal}}}, \bibinfo {author} {\bibfnamefont {L.~N.}\ \bibnamefont {{da
  Costa}}}, \bibinfo {author} {\bibfnamefont {D.~S.}\ \bibnamefont
  {{Goldwirth}}}, \bibinfo {author} {\bibfnamefont {M.}~\bibnamefont
  {{Lecar}}}, \ and\ \bibinfo {author} {\bibfnamefont {T.}~\bibnamefont
  {{Piran}}},\ }\href {\doibase 10.1086/171147} {\bibfield  {journal} {\bibinfo
   {journal} {\apj}\ }\textbf {\bibinfo {volume} {388}},\ \bibinfo {pages}
  {234} (\bibinfo {year} {1992})}\BibitemShut {NoStop}%
\bibitem [{\citenamefont {{Dubinski}}\ \emph {et~al.}(1993)\citenamefont
  {{Dubinski}}, \citenamefont {{da Costa}}, \citenamefont {{Goldwirth}},
  \citenamefont {{Lecar}},\ and\ \citenamefont {{Piran}}}]{DCG93}%
  \BibitemOpen
  \bibfield  {author} {\bibinfo {author} {\bibfnamefont {J.}~\bibnamefont
  {{Dubinski}}}, \bibinfo {author} {\bibfnamefont {L.~N.}\ \bibnamefont {{da
  Costa}}}, \bibinfo {author} {\bibfnamefont {D.~S.}\ \bibnamefont
  {{Goldwirth}}}, \bibinfo {author} {\bibfnamefont {M.}~\bibnamefont
  {{Lecar}}}, \ and\ \bibinfo {author} {\bibfnamefont {T.}~\bibnamefont
  {{Piran}}},\ }\href {\doibase 10.1086/172762} {\bibfield  {journal} {\bibinfo
   {journal} {\apj}\ }\textbf {\bibinfo {volume} {410}},\ \bibinfo {pages}
  {458} (\bibinfo {year} {1993})}\BibitemShut {NoStop}%
\bibitem [{\citenamefont {{van de Weygaert}}\ and\ \citenamefont {{van
  Kampen}}(1993)}]{WK93}%
  \BibitemOpen
  \bibfield  {author} {\bibinfo {author} {\bibfnamefont {R.}~\bibnamefont {{van
  de Weygaert}}}\ and\ \bibinfo {author} {\bibfnamefont {E.}~\bibnamefont {{van
  Kampen}}},\ }\href@noop {} {\bibfield  {journal} {\bibinfo  {journal}
  {\mnras}\ }\textbf {\bibinfo {volume} {263}},\ \bibinfo {pages} {481}
  (\bibinfo {year} {1993})}\BibitemShut {NoStop}%
\bibitem [{\citenamefont {{Gottl{\"o}ber}}\ \emph {et~al.}(2003)\citenamefont
  {{Gottl{\"o}ber}}, \citenamefont {{{\L}okas}}, \citenamefont {{Klypin}},\
  and\ \citenamefont {{Hoffman}}}]{GLK03}%
  \BibitemOpen
  \bibfield  {author} {\bibinfo {author} {\bibfnamefont {S.}~\bibnamefont
  {{Gottl{\"o}ber}}}, \bibinfo {author} {\bibfnamefont {E.~L.}\ \bibnamefont
  {{{\L}okas}}}, \bibinfo {author} {\bibfnamefont {A.}~\bibnamefont
  {{Klypin}}}, \ and\ \bibinfo {author} {\bibfnamefont {Y.}~\bibnamefont
  {{Hoffman}}},\ }\href {\doibase 10.1046/j.1365-8711.2003.06850.x} {\bibfield
  {journal} {\bibinfo  {journal} {\mnras}\ }\textbf {\bibinfo {volume} {344}},\
  \bibinfo {pages} {715} (\bibinfo {year} {2003})},\ \Eprint
  {http://arxiv.org/abs/astro-ph/0305393} {astro-ph/0305393} \BibitemShut
  {NoStop}%
\bibitem [{\citenamefont {{Colberg}}\ \emph {et~al.}(2005)\citenamefont
  {{Colberg}}, \citenamefont {{Sheth}}, \citenamefont {{Diaferio}},
  \citenamefont {{Gao}},\ and\ \citenamefont {{Yoshida}}}]{CSD05}%
  \BibitemOpen
  \bibfield  {author} {\bibinfo {author} {\bibfnamefont {J.~M.}\ \bibnamefont
  {{Colberg}}}, \bibinfo {author} {\bibfnamefont {R.~K.}\ \bibnamefont
  {{Sheth}}}, \bibinfo {author} {\bibfnamefont {A.}~\bibnamefont {{Diaferio}}},
  \bibinfo {author} {\bibfnamefont {L.}~\bibnamefont {{Gao}}}, \ and\ \bibinfo
  {author} {\bibfnamefont {N.}~\bibnamefont {{Yoshida}}},\ }\href {\doibase
  10.1111/j.1365-2966.2005.09064.x} {\bibfield  {journal} {\bibinfo  {journal}
  {\mnras}\ }\textbf {\bibinfo {volume} {360}},\ \bibinfo {pages} {216}
  (\bibinfo {year} {2005})},\ \Eprint {http://arxiv.org/abs/astro-ph/0409162}
  {astro-ph/0409162} \BibitemShut {NoStop}%
\bibitem [{\citenamefont {{Platen}}\ \emph {et~al.}(2008)\citenamefont
  {{Platen}}, \citenamefont {{van de Weygaert}},\ and\ \citenamefont
  {{Jones}}}]{PWJ08}%
  \BibitemOpen
  \bibfield  {author} {\bibinfo {author} {\bibfnamefont {E.}~\bibnamefont
  {{Platen}}}, \bibinfo {author} {\bibfnamefont {R.}~\bibnamefont {{van de
  Weygaert}}}, \ and\ \bibinfo {author} {\bibfnamefont {B.~J.~T.}\ \bibnamefont
  {{Jones}}},\ }\href {\doibase 10.1111/j.1365-2966.2008.13019.x} {\bibfield
  {journal} {\bibinfo  {journal} {\mnras}\ }\textbf {\bibinfo {volume} {387}},\
  \bibinfo {pages} {128} (\bibinfo {year} {2008})},\ \Eprint
  {http://arxiv.org/abs/0711.2480} {arXiv:0711.2480} \BibitemShut {NoStop}%
\bibitem [{\citenamefont {{Einasto}}\ \emph {et~al.}(2011)\citenamefont
  {{Einasto}}, \citenamefont {{Suhhonenko}}, \citenamefont {{H{\"u}tsi}},
  \citenamefont {{Saar}}, \citenamefont {{Einasto}}, \citenamefont
  {{Liivam{\"a}gi}}, \citenamefont {{M{\"u}ller}}, \citenamefont
  {{Starobinsky}},\ and\ \citenamefont {et~al.}}]{ESH11}%
  \BibitemOpen
  \bibfield  {author} {\bibinfo {author} {\bibfnamefont {J.}~\bibnamefont
  {{Einasto}}}, \bibinfo {author} {\bibfnamefont {I.}~\bibnamefont
  {{Suhhonenko}}}, \bibinfo {author} {\bibfnamefont {G.}~\bibnamefont
  {{H{\"u}tsi}}}, \bibinfo {author} {\bibfnamefont {E.}~\bibnamefont {{Saar}}},
  \bibinfo {author} {\bibfnamefont {M.}~\bibnamefont {{Einasto}}}, \bibinfo
  {author} {\bibfnamefont {L.~J.}\ \bibnamefont {{Liivam{\"a}gi}}}, \bibinfo
  {author} {\bibfnamefont {V.}~\bibnamefont {{M{\"u}ller}}}, \bibinfo {author}
  {\bibfnamefont {A.~A.}\ \bibnamefont {{Starobinsky}}}, \ and\ \bibinfo
  {author} {\bibnamefont {et~al.}},\ }\href {\doibase
  10.1051/0004-6361/201117248} {\bibfield  {journal} {\bibinfo  {journal}
  {\aap}\ }\textbf {\bibinfo {volume} {534}},\ \bibinfo {eid} {A128} (\bibinfo
  {year} {2011})},\ \Eprint {http://arxiv.org/abs/1105.2464} {arXiv:1105.2464
  [astro-ph.CO]} \BibitemShut {NoStop}%
\bibitem [{\citenamefont {{Aikio}}\ and\ \citenamefont
  {{M{\"a}h{\"o}nen}}(1998)}]{AM98}%
  \BibitemOpen
  \bibfield  {author} {\bibinfo {author} {\bibfnamefont {J.}~\bibnamefont
  {{Aikio}}}\ and\ \bibinfo {author} {\bibfnamefont {P.}~\bibnamefont
  {{M{\"a}h{\"o}nen}}},\ }\href {\doibase 10.1086/305509} {\bibfield  {journal}
  {\bibinfo  {journal} {\apj}\ }\textbf {\bibinfo {volume} {497}},\ \bibinfo
  {pages} {534} (\bibinfo {year} {1998})}\BibitemShut {NoStop}%
\bibitem [{\citenamefont {{Friedmann}}\ and\ \citenamefont
  {{Piran}}(2001)}]{FP01}%
  \BibitemOpen
  \bibfield  {author} {\bibinfo {author} {\bibfnamefont {Y.}~\bibnamefont
  {{Friedmann}}}\ and\ \bibinfo {author} {\bibfnamefont {T.}~\bibnamefont
  {{Piran}}},\ }\href {\doibase 10.1086/318652} {\bibfield  {journal} {\bibinfo
   {journal} {\apj}\ }\textbf {\bibinfo {volume} {548}},\ \bibinfo {pages} {1}
  (\bibinfo {year} {2001})},\ \Eprint {http://arxiv.org/abs/astro-ph/0009320}
  {astro-ph/0009320} \BibitemShut {NoStop}%
\bibitem [{\citenamefont {{Neyrinck}}\ \emph {et~al.}(2005)\citenamefont
  {{Neyrinck}}, \citenamefont {{Gnedin}},\ and\ \citenamefont
  {{Hamilton}}}]{Netal05}%
  \BibitemOpen
  \bibfield  {author} {\bibinfo {author} {\bibfnamefont {M.~C.}\ \bibnamefont
  {{Neyrinck}}}, \bibinfo {author} {\bibfnamefont {N.~Y.}\ \bibnamefont
  {{Gnedin}}}, \ and\ \bibinfo {author} {\bibfnamefont {A.~J.~S.}\ \bibnamefont
  {{Hamilton}}},\ }\href {\doibase 10.1111/j.1365-2966.2004.08505.x} {\bibfield
   {journal} {\bibinfo  {journal} {\mnras}\ }\textbf {\bibinfo {volume}
  {356}},\ \bibinfo {pages} {1222} (\bibinfo {year} {2005})},\ \Eprint
  {http://arxiv.org/abs/astro-ph/0402346} {astro-ph/0402346} \BibitemShut
  {NoStop}%
\bibitem [{\citenamefont {{Gaite}}(2005)}]{2005EPJB...47...93G}%
  \BibitemOpen
  \bibfield  {author} {\bibinfo {author} {\bibfnamefont {J.}~\bibnamefont
  {{Gaite}}},\ }\href {\doibase 10.1140/epjb/e2005-00306-1} {\bibfield
  {journal} {\bibinfo  {journal} {European Physical Journal B}\ }\textbf
  {\bibinfo {volume} {47}},\ \bibinfo {pages} {93} (\bibinfo {year} {2005})},\
  \Eprint {http://arxiv.org/abs/astro-ph/0506543} {astro-ph/0506543}
  \BibitemShut {NoStop}%
\bibitem [{\citenamefont {{Patiri}}\ \emph
  {et~al.}(2006{\natexlab{b}})\citenamefont {{Patiri}}, \citenamefont
  {{Prada}}, \citenamefont {{Holtzman}}, \citenamefont {{Klypin}},\ and\
  \citenamefont {{Betancort-Rijo}}}]{PPH06}%
  \BibitemOpen
  \bibfield  {author} {\bibinfo {author} {\bibfnamefont {S.~G.}\ \bibnamefont
  {{Patiri}}}, \bibinfo {author} {\bibfnamefont {F.}~\bibnamefont {{Prada}}},
  \bibinfo {author} {\bibfnamefont {J.}~\bibnamefont {{Holtzman}}}, \bibinfo
  {author} {\bibfnamefont {A.}~\bibnamefont {{Klypin}}}, \ and\ \bibinfo
  {author} {\bibfnamefont {J.}~\bibnamefont {{Betancort-Rijo}}},\ }\href
  {\doibase 10.1111/j.1365-2966.2006.10975.x} {\bibfield  {journal} {\bibinfo
  {journal} {\mnras}\ }\textbf {\bibinfo {volume} {372}},\ \bibinfo {pages}
  {1710} (\bibinfo {year} {2006}{\natexlab{b}})},\ \Eprint
  {http://arxiv.org/abs/astro-ph/0605703} {astro-ph/0605703} \BibitemShut
  {NoStop}%
\bibitem [{\citenamefont {{Platen}}\ \emph {et~al.}(2007)\citenamefont
  {{Platen}}, \citenamefont {{van de Weygaert}},\ and\ \citenamefont
  {{Jones}}}]{PWJ07}%
  \BibitemOpen
  \bibfield  {author} {\bibinfo {author} {\bibfnamefont {E.}~\bibnamefont
  {{Platen}}}, \bibinfo {author} {\bibfnamefont {R.}~\bibnamefont {{van de
  Weygaert}}}, \ and\ \bibinfo {author} {\bibfnamefont {B.~J.~T.}\ \bibnamefont
  {{Jones}}},\ }\href {\doibase 10.1111/j.1365-2966.2007.12125.x} {\bibfield
  {journal} {\bibinfo  {journal} {\mnras}\ }\textbf {\bibinfo {volume} {380}},\
  \bibinfo {pages} {551} (\bibinfo {year} {2007})},\ \Eprint
  {http://arxiv.org/abs/0706.2788} {arXiv:0706.2788} \BibitemShut {NoStop}%
\bibitem [{\citenamefont {{Shandarin}}\ \emph {et~al.}(2012)\citenamefont
  {{Shandarin}}, \citenamefont {{Habib}},\ and\ \citenamefont
  {{Heitmann}}}]{SHH12}%
  \BibitemOpen
  \bibfield  {author} {\bibinfo {author} {\bibfnamefont {S.}~\bibnamefont
  {{Shandarin}}}, \bibinfo {author} {\bibfnamefont {S.}~\bibnamefont
  {{Habib}}}, \ and\ \bibinfo {author} {\bibfnamefont {K.}~\bibnamefont
  {{Heitmann}}},\ }\href {\doibase 10.1103/PhysRevD.85.083005} {\bibfield
  {journal} {\bibinfo  {journal} {\prd}\ }\textbf {\bibinfo {volume} {85}},\
  \bibinfo {eid} {083005} (\bibinfo {year} {2012})},\ \Eprint
  {http://arxiv.org/abs/1111.2366} {arXiv:1111.2366} \BibitemShut {NoStop}%
\bibitem [{\citenamefont {{Abel}}\ \emph {et~al.}(2012)\citenamefont {{Abel}},
  \citenamefont {{Hahn}},\ and\ \citenamefont {{Kaehler}}}]{AHK12}%
  \BibitemOpen
  \bibfield  {author} {\bibinfo {author} {\bibfnamefont {T.}~\bibnamefont
  {{Abel}}}, \bibinfo {author} {\bibfnamefont {O.}~\bibnamefont {{Hahn}}}, \
  and\ \bibinfo {author} {\bibfnamefont {R.}~\bibnamefont {{Kaehler}}},\ }\href
  {\doibase 10.1111/j.1365-2966.2012.21754.x} {\bibfield  {journal} {\bibinfo
  {journal} {\mnras}\ }\textbf {\bibinfo {volume} {427}},\ \bibinfo {pages}
  {61} (\bibinfo {year} {2012})},\ \Eprint {http://arxiv.org/abs/1111.3944}
  {arXiv:1111.3944} \BibitemShut {NoStop}%
\bibitem [{\citenamefont {{Falck}}\ \emph {et~al.}(2012)\citenamefont
  {{Falck}}, \citenamefont {{Neyrinck}},\ and\ \citenamefont
  {{Szalay}}}]{FNS12}%
  \BibitemOpen
  \bibfield  {author} {\bibinfo {author} {\bibfnamefont {B.~L.}\ \bibnamefont
  {{Falck}}}, \bibinfo {author} {\bibfnamefont {M.~C.}\ \bibnamefont
  {{Neyrinck}}}, \ and\ \bibinfo {author} {\bibfnamefont {A.~S.}\ \bibnamefont
  {{Szalay}}},\ }\href {\doibase 10.1088/0004-637X/754/2/126} {\bibfield
  {journal} {\bibinfo  {journal} {\apj}\ }\textbf {\bibinfo {volume} {754}},\
  \bibinfo {eid} {126} (\bibinfo {year} {2012})},\ \Eprint
  {http://arxiv.org/abs/1201.2353} {arXiv:1201.2353} \BibitemShut {NoStop}%
\bibitem [{\citenamefont {{Cautun}}\ \emph {et~al.}(2013)\citenamefont
  {{Cautun}}, \citenamefont {{van de Weygaert}},\ and\ \citenamefont
  {{Jones}}}]{CWJ13}%
  \BibitemOpen
  \bibfield  {author} {\bibinfo {author} {\bibfnamefont {M.}~\bibnamefont
  {{Cautun}}}, \bibinfo {author} {\bibfnamefont {R.}~\bibnamefont {{van de
  Weygaert}}}, \ and\ \bibinfo {author} {\bibfnamefont {B.~J.~T.}\ \bibnamefont
  {{Jones}}},\ }\href {\doibase 10.1093/mnras/sts416} {\bibfield  {journal}
  {\bibinfo  {journal} {\mnras}\ }\textbf {\bibinfo {volume} {429}},\ \bibinfo
  {pages} {1286} (\bibinfo {year} {2013})},\ \Eprint
  {http://arxiv.org/abs/1209.2043} {arXiv:1209.2043} \BibitemShut {NoStop}%
\bibitem [{\citenamefont {{Way}}\ \emph {et~al.}(2015)\citenamefont {{Way}},
  \citenamefont {{Gazis}},\ and\ \citenamefont
  {{Scargle}}}]{2015ApJ...799...95W}%
  \BibitemOpen
  \bibfield  {author} {\bibinfo {author} {\bibfnamefont {M.~J.}\ \bibnamefont
  {{Way}}}, \bibinfo {author} {\bibfnamefont {P.~R.}\ \bibnamefont {{Gazis}}},
  \ and\ \bibinfo {author} {\bibfnamefont {J.~D.}\ \bibnamefont {{Scargle}}},\
  }\href {\doibase 10.1088/0004-637X/799/1/95} {\bibfield  {journal} {\bibinfo
  {journal} {\apj}\ }\textbf {\bibinfo {volume} {799}},\ \bibinfo {eid} {95}
  (\bibinfo {year} {2015})},\ \Eprint {http://arxiv.org/abs/1406.6111}
  {arXiv:1406.6111} \BibitemShut {NoStop}%
\bibitem [{\citenamefont {{Colberg et al.}}(2008)}]{CPF08}%
  \BibitemOpen
  \bibfield  {author} {\bibinfo {author} {\bibfnamefont {J.~M.}\ \bibnamefont
  {{Colberg et al.}}},\ }\href {\doibase 10.1111/j.1365-2966.2008.13307.x}
  {\bibfield  {journal} {\bibinfo  {journal} {\mnras}\ }\textbf {\bibinfo
  {volume} {387}},\ \bibinfo {pages} {933} (\bibinfo {year} {2008})},\ \Eprint
  {http://arxiv.org/abs/0803.0918} {arXiv:0803.0918} \BibitemShut {NoStop}%
\bibitem [{\citenamefont {{Park}}\ and\ \citenamefont {{Lee}}(2007)}]{PL07}%
  \BibitemOpen
  \bibfield  {author} {\bibinfo {author} {\bibfnamefont {D.}~\bibnamefont
  {{Park}}}\ and\ \bibinfo {author} {\bibfnamefont {J.}~\bibnamefont {{Lee}}},\
  }\href {\doibase 10.1103/PhysRevLett.98.081301} {\bibfield  {journal}
  {\bibinfo  {journal} {Physical Review Letters}\ }\textbf {\bibinfo {volume}
  {98}},\ \bibinfo {eid} {081301} (\bibinfo {year} {2007})},\ \Eprint
  {http://arxiv.org/abs/astro-ph/0610520} {astro-ph/0610520} \BibitemShut
  {NoStop}%
\bibitem [{\citenamefont {{Lavaux}}\ and\ \citenamefont
  {{Wandelt}}(2010)}]{LW10}%
  \BibitemOpen
  \bibfield  {author} {\bibinfo {author} {\bibfnamefont {G.}~\bibnamefont
  {{Lavaux}}}\ and\ \bibinfo {author} {\bibfnamefont {B.~D.}\ \bibnamefont
  {{Wandelt}}},\ }\href {\doibase 10.1111/j.1365-2966.2010.16197.x} {\bibfield
  {journal} {\bibinfo  {journal} {\mnras}\ }\textbf {\bibinfo {volume} {403}},\
  \bibinfo {pages} {1392} (\bibinfo {year} {2010})},\ \Eprint
  {http://arxiv.org/abs/0906.4101} {arXiv:0906.4101} \BibitemShut {NoStop}%
\bibitem [{\citenamefont {{Bos}}\ \emph {et~al.}(2012)\citenamefont {{Bos}},
  \citenamefont {{van de Weygaert}}, \citenamefont {{Dolag}},\ and\
  \citenamefont {{Pettorino}}}]{B12}%
  \BibitemOpen
  \bibfield  {author} {\bibinfo {author} {\bibfnamefont {E.~G.~P.}\
  \bibnamefont {{Bos}}}, \bibinfo {author} {\bibfnamefont {R.}~\bibnamefont
  {{van de Weygaert}}}, \bibinfo {author} {\bibfnamefont {K.}~\bibnamefont
  {{Dolag}}}, \ and\ \bibinfo {author} {\bibfnamefont {V.}~\bibnamefont
  {{Pettorino}}},\ }\href {\doibase 10.1111/j.1365-2966.2012.21478.x}
  {\bibfield  {journal} {\bibinfo  {journal} {\mnras}\ }\textbf {\bibinfo
  {volume} {426}},\ \bibinfo {pages} {440} (\bibinfo {year} {2012})},\ \Eprint
  {http://arxiv.org/abs/1205.4238} {arXiv:1205.4238 [astro-ph.CO]} \BibitemShut
  {NoStop}%
\bibitem [{\citenamefont {{Pisani}}\ \emph {et~al.}(2015)\citenamefont
  {{Pisani}}, \citenamefont {{Sutter}}, \citenamefont {{Hamaus}}, \citenamefont
  {{Alizadeh}}, \citenamefont {{Biswas}}, \citenamefont {{Wandelt}},\ and\
  \citenamefont {{Hirata}}}]{2015arXiv150307690P}%
  \BibitemOpen
  \bibfield  {author} {\bibinfo {author} {\bibfnamefont {A.}~\bibnamefont
  {{Pisani}}}, \bibinfo {author} {\bibfnamefont {P.~M.}\ \bibnamefont
  {{Sutter}}}, \bibinfo {author} {\bibfnamefont {N.}~\bibnamefont {{Hamaus}}},
  \bibinfo {author} {\bibfnamefont {E.}~\bibnamefont {{Alizadeh}}}, \bibinfo
  {author} {\bibfnamefont {R.}~\bibnamefont {{Biswas}}}, \bibinfo {author}
  {\bibfnamefont {B.~D.}\ \bibnamefont {{Wandelt}}}, \ and\ \bibinfo {author}
  {\bibfnamefont {C.~M.}\ \bibnamefont {{Hirata}}},\ }\href@noop {} {\bibfield
  {journal} {\bibinfo  {journal} {ArXiv e-prints}\ } (\bibinfo {year}
  {2015})},\ \Eprint {http://arxiv.org/abs/1503.07690} {arXiv:1503.07690}
  \BibitemShut {NoStop}%
\bibitem [{\citenamefont {{Sutter}}\ \emph
  {et~al.}(2014{\natexlab{b}})\citenamefont {{Sutter}}, \citenamefont
  {{Pisani}}, \citenamefont {{Wandelt}},\ and\ \citenamefont
  {{Weinberg}}}]{SPW14}%
  \BibitemOpen
  \bibfield  {author} {\bibinfo {author} {\bibfnamefont {P.~M.}\ \bibnamefont
  {{Sutter}}}, \bibinfo {author} {\bibfnamefont {A.}~\bibnamefont {{Pisani}}},
  \bibinfo {author} {\bibfnamefont {B.~D.}\ \bibnamefont {{Wandelt}}}, \ and\
  \bibinfo {author} {\bibfnamefont {D.~H.}\ \bibnamefont {{Weinberg}}},\ }\href
  {\doibase 10.1093/mnras/stu1392} {\bibfield  {journal} {\bibinfo  {journal}
  {\mnras}\ }\textbf {\bibinfo {volume} {443}},\ \bibinfo {pages} {2983}
  (\bibinfo {year} {2014}{\natexlab{b}})},\ \Eprint
  {http://arxiv.org/abs/1404.5618} {arXiv:1404.5618} \BibitemShut {NoStop}%
\bibitem [{\citenamefont {{Lam}}\ \emph {et~al.}(2015)\citenamefont {{Lam}},
  \citenamefont {{Clampitt}}, \citenamefont {{Cai}},\ and\ \citenamefont
  {{Li}}}]{LCC15}%
  \BibitemOpen
  \bibfield  {author} {\bibinfo {author} {\bibfnamefont {T.~Y.}\ \bibnamefont
  {{Lam}}}, \bibinfo {author} {\bibfnamefont {J.}~\bibnamefont {{Clampitt}}},
  \bibinfo {author} {\bibfnamefont {Y.-C.}\ \bibnamefont {{Cai}}}, \ and\
  \bibinfo {author} {\bibfnamefont {B.}~\bibnamefont {{Li}}},\ }\href {\doibase
  10.1093/mnras/stv797} {\bibfield  {journal} {\bibinfo  {journal} {\mnras}\
  }\textbf {\bibinfo {volume} {450}},\ \bibinfo {pages} {3319} (\bibinfo {year}
  {2015})},\ \Eprint {http://arxiv.org/abs/1408.5338} {arXiv:1408.5338}
  \BibitemShut {NoStop}%
\bibitem [{\citenamefont {{Cai}}\ \emph {et~al.}(2015)\citenamefont {{Cai}},
  \citenamefont {{Padilla}},\ and\ \citenamefont {{Li}}}]{C15}%
  \BibitemOpen
  \bibfield  {author} {\bibinfo {author} {\bibfnamefont {Y.-C.}\ \bibnamefont
  {{Cai}}}, \bibinfo {author} {\bibfnamefont {N.}~\bibnamefont {{Padilla}}}, \
  and\ \bibinfo {author} {\bibfnamefont {B.}~\bibnamefont {{Li}}},\ }\href
  {\doibase 10.1093/mnras/stv777} {\bibfield  {journal} {\bibinfo  {journal}
  {\mnras}\ }\textbf {\bibinfo {volume} {451}},\ \bibinfo {pages} {5555}
  (\bibinfo {year} {2015})},\ \Eprint {http://arxiv.org/abs/1410.1510}
  {arXiv:1410.1510} \BibitemShut {NoStop}%
\bibitem [{\citenamefont {{Li}}(2011)}]{L11}%
  \BibitemOpen
  \bibfield  {author} {\bibinfo {author} {\bibfnamefont {B.}~\bibnamefont
  {{Li}}},\ }\href {\doibase 10.1111/j.1365-2966.2010.17867.x} {\bibfield
  {journal} {\bibinfo  {journal} {\mnras}\ }\textbf {\bibinfo {volume} {411}},\
  \bibinfo {pages} {2615} (\bibinfo {year} {2011})},\ \Eprint
  {http://arxiv.org/abs/1009.1406} {arXiv:1009.1406 [astro-ph.CO]} \BibitemShut
  {NoStop}%
\bibitem [{\citenamefont {{Martino}}\ and\ \citenamefont
  {{Sheth}}(2009)}]{MS09}%
  \BibitemOpen
  \bibfield  {author} {\bibinfo {author} {\bibfnamefont {M.~C.}\ \bibnamefont
  {{Martino}}}\ and\ \bibinfo {author} {\bibfnamefont {R.~K.}\ \bibnamefont
  {{Sheth}}},\ }\href@noop {} {\bibfield  {journal} {\bibinfo  {journal} {ArXiv
  e-prints}\ } (\bibinfo {year} {2009})},\ \Eprint
  {http://arxiv.org/abs/0911.1829} {arXiv:0911.1829 [astro-ph.CO]} \BibitemShut
  {NoStop}%
\bibitem [{\citenamefont {{Planck
  Collaboration}}(2014{\natexlab{b}})}]{PLANCKISW14}%
  \BibitemOpen
  \bibfield  {author} {\bibinfo {author} {\bibnamefont {{Planck
  Collaboration}}},\ }\href {\doibase 10.1051/0004-6361/201321526} {\bibfield
  {journal} {\bibinfo  {journal} {\aap}\ }\textbf {\bibinfo {volume} {571}},\
  \bibinfo {eid} {A19} (\bibinfo {year} {2014}{\natexlab{b}})},\ \Eprint
  {http://arxiv.org/abs/1303.5079} {arXiv:1303.5079} \BibitemShut {NoStop}%
\bibitem [{\citenamefont {{Goldwirth}}\ \emph {et~al.}(1995)\citenamefont
  {{Goldwirth}}, \citenamefont {{da Costa}},\ and\ \citenamefont {{van de
  Weygaert}}}]{GCW95}%
  \BibitemOpen
  \bibfield  {author} {\bibinfo {author} {\bibfnamefont {D.~S.}\ \bibnamefont
  {{Goldwirth}}}, \bibinfo {author} {\bibfnamefont {L.~N.}\ \bibnamefont {{da
  Costa}}}, \ and\ \bibinfo {author} {\bibfnamefont {R.}~\bibnamefont {{van de
  Weygaert}}},\ }\href@noop {} {\bibfield  {journal} {\bibinfo  {journal}
  {\mnras}\ }\textbf {\bibinfo {volume} {275}},\ \bibinfo {pages} {1185}
  (\bibinfo {year} {1995})},\ \Eprint {http://arxiv.org/abs/astro-ph/9503002}
  {astro-ph/9503002} \BibitemShut {NoStop}%
\bibitem [{\citenamefont {{Padilla}}\ \emph {et~al.}(2005)\citenamefont
  {{Padilla}}, \citenamefont {{Ceccarelli}},\ and\ \citenamefont
  {{Lambas}}}]{padilla05}%
  \BibitemOpen
  \bibfield  {author} {\bibinfo {author} {\bibfnamefont {N.~D.}\ \bibnamefont
  {{Padilla}}}, \bibinfo {author} {\bibfnamefont {L.}~\bibnamefont
  {{Ceccarelli}}}, \ and\ \bibinfo {author} {\bibfnamefont {D.~G.}\
  \bibnamefont {{Lambas}}},\ }\href {\doibase 10.1111/j.1365-2966.2005.09500.x}
  {\bibfield  {journal} {\bibinfo  {journal} {\mnras}\ }\textbf {\bibinfo
  {volume} {363}},\ \bibinfo {pages} {977} (\bibinfo {year} {2005})},\ \Eprint
  {http://arxiv.org/abs/astro-ph/0508297} {astro-ph/0508297} \BibitemShut
  {NoStop}%
\bibitem [{\citenamefont {{Hamaus}}\ \emph {et~al.}(2014)\citenamefont
  {{Hamaus}}, \citenamefont {{Sutter}},\ and\ \citenamefont
  {{Wandelt}}}]{HSW14}%
  \BibitemOpen
  \bibfield  {author} {\bibinfo {author} {\bibfnamefont {N.}~\bibnamefont
  {{Hamaus}}}, \bibinfo {author} {\bibfnamefont {P.~M.}\ \bibnamefont
  {{Sutter}}}, \ and\ \bibinfo {author} {\bibfnamefont {B.~D.}\ \bibnamefont
  {{Wandelt}}},\ }\href {\doibase 10.1103/PhysRevLett.112.251302} {\bibfield
  {journal} {\bibinfo  {journal} {Physical Review Letters}\ }\textbf {\bibinfo
  {volume} {112}},\ \bibinfo {eid} {251302} (\bibinfo {year} {2014})},\ \Eprint
  {http://arxiv.org/abs/1403.5499} {arXiv:1403.5499} \BibitemShut {NoStop}%
\bibitem [{\citenamefont {{Clampitt}}\ \emph {et~al.}(2015)\citenamefont
  {{Clampitt}}, \citenamefont {{Jain}},\ and\ \citenamefont
  {{S{\'a}nchez}}}]{CJS15}%
  \BibitemOpen
  \bibfield  {author} {\bibinfo {author} {\bibfnamefont {J.}~\bibnamefont
  {{Clampitt}}}, \bibinfo {author} {\bibfnamefont {B.}~\bibnamefont {{Jain}}},
  \ and\ \bibinfo {author} {\bibfnamefont {C.}~\bibnamefont {{S{\'a}nchez}}},\
  }\href@noop {} {\bibfield  {journal} {\bibinfo  {journal} {ArXiv e-prints}\ }
  (\bibinfo {year} {2015})},\ \Eprint {http://arxiv.org/abs/1507.08031}
  {arXiv:1507.08031} \BibitemShut {NoStop}%
\bibitem [{\citenamefont {{Zhao}}\ \emph
  {et~al.}(2015{\natexlab{a}})\citenamefont {{Zhao}}, \citenamefont {{Tao}},
  \citenamefont {{Liang}}, \citenamefont {{Kitaura}},\ and\ \citenamefont
  {{Chuang}}}]{2015arXiv151104299Z}%
  \BibitemOpen
  \bibfield  {author} {\bibinfo {author} {\bibfnamefont {C.}~\bibnamefont
  {{Zhao}}}, \bibinfo {author} {\bibfnamefont {C.}~\bibnamefont {{Tao}}},
  \bibinfo {author} {\bibfnamefont {Y.}~\bibnamefont {{Liang}}}, \bibinfo
  {author} {\bibfnamefont {F.-S.}\ \bibnamefont {{Kitaura}}}, \ and\ \bibinfo
  {author} {\bibfnamefont {C.-H.}\ \bibnamefont {{Chuang}}},\ }\href@noop {}
  {\bibfield  {journal} {\bibinfo  {journal} {ArXiv e-prints}\ } (\bibinfo
  {year} {2015}{\natexlab{a}})},\ \Eprint {http://arxiv.org/abs/1511.04299}
  {arXiv:1511.04299} \BibitemShut {NoStop}%
\bibitem [{\citenamefont {{Liang}}\ \emph {et~al.}(2015)\citenamefont
  {{Liang}}, \citenamefont {{Zhao}}, \citenamefont {{Chuang}}, \citenamefont
  {{Kitaura}},\ and\ \citenamefont {{Tao}}}]{2015arXiv151104391L}%
  \BibitemOpen
  \bibfield  {author} {\bibinfo {author} {\bibfnamefont {Y.}~\bibnamefont
  {{Liang}}}, \bibinfo {author} {\bibfnamefont {C.}~\bibnamefont {{Zhao}}},
  \bibinfo {author} {\bibfnamefont {C.-H.}\ \bibnamefont {{Chuang}}}, \bibinfo
  {author} {\bibfnamefont {F.-S.}\ \bibnamefont {{Kitaura}}}, \ and\ \bibinfo
  {author} {\bibfnamefont {C.}~\bibnamefont {{Tao}}},\ }\href@noop {}
  {\bibfield  {journal} {\bibinfo  {journal} {ArXiv e-prints}\ } (\bibinfo
  {year} {2015})},\ \Eprint {http://arxiv.org/abs/1511.04391}
  {arXiv:1511.04391} \BibitemShut {NoStop}%
\bibitem [{\citenamefont {{Kitaura}}\ \emph {et~al.}(2014)\citenamefont
  {{Kitaura}}, \citenamefont {{Yepes}},\ and\ \citenamefont
  {{Prada}}}]{KitauraPatchy}%
  \BibitemOpen
  \bibfield  {author} {\bibinfo {author} {\bibfnamefont {F.-S.}\ \bibnamefont
  {{Kitaura}}}, \bibinfo {author} {\bibfnamefont {G.}~\bibnamefont {{Yepes}}},
  \ and\ \bibinfo {author} {\bibfnamefont {F.}~\bibnamefont {{Prada}}},\ }\href
  {\doibase 10.1093/mnrasl/slt172} {\bibfield  {journal} {\bibinfo  {journal}
  {\mnras}\ }\textbf {\bibinfo {volume} {439}},\ \bibinfo {pages} {L21}
  (\bibinfo {year} {2014})},\ \Eprint {http://arxiv.org/abs/1307.3285}
  {arXiv:1307.3285 [astro-ph.CO]} \BibitemShut {NoStop}%
\bibitem [{\citenamefont {{Klypin}}\ \emph {et~al.}(2014)\citenamefont
  {{Klypin}}, \citenamefont {{Yepes}}, \citenamefont {{Gottlober}},
  \citenamefont {{Prada}},\ and\ \citenamefont {{Hess}}}]{Klypin2014}%
  \BibitemOpen
  \bibfield  {author} {\bibinfo {author} {\bibfnamefont {A.}~\bibnamefont
  {{Klypin}}}, \bibinfo {author} {\bibfnamefont {G.}~\bibnamefont {{Yepes}}},
  \bibinfo {author} {\bibfnamefont {S.}~\bibnamefont {{Gottlober}}}, \bibinfo
  {author} {\bibfnamefont {F.}~\bibnamefont {{Prada}}}, \ and\ \bibinfo
  {author} {\bibfnamefont {S.}~\bibnamefont {{Hess}}},\ }\href@noop {}
  {\bibfield  {journal} {\bibinfo  {journal} {ArXiv e-prints}\ } (\bibinfo
  {year} {2014})},\ \Eprint {http://arxiv.org/abs/1411.4001} {arXiv:1411.4001}
  \BibitemShut {NoStop}%
\bibitem [{\citenamefont {{Kitaura}}\ \emph
  {et~al.}(2015{\natexlab{a}})\citenamefont {{Kitaura}}, \citenamefont
  {{Gil-Mar{\'{\i}}n}}, \citenamefont {{Sc{\'o}ccola}}, \citenamefont
  {{Chuang}}, \citenamefont {{M{\"u}ller}}, \citenamefont {{Yepes}},\ and\
  \citenamefont {{Prada}}}]{KGS15}%
  \BibitemOpen
  \bibfield  {author} {\bibinfo {author} {\bibfnamefont {F.-S.}\ \bibnamefont
  {{Kitaura}}}, \bibinfo {author} {\bibfnamefont {H.}~\bibnamefont
  {{Gil-Mar{\'{\i}}n}}}, \bibinfo {author} {\bibfnamefont {C.~G.}\ \bibnamefont
  {{Sc{\'o}ccola}}}, \bibinfo {author} {\bibfnamefont {C.-H.}\ \bibnamefont
  {{Chuang}}}, \bibinfo {author} {\bibfnamefont {V.}~\bibnamefont
  {{M{\"u}ller}}}, \bibinfo {author} {\bibfnamefont {G.}~\bibnamefont
  {{Yepes}}}, \ and\ \bibinfo {author} {\bibfnamefont {F.}~\bibnamefont
  {{Prada}}},\ }\href {\doibase 10.1093/mnras/stv645} {\bibfield  {journal}
  {\bibinfo  {journal} {\mnras}\ }\textbf {\bibinfo {volume} {450}},\ \bibinfo
  {pages} {1836} (\bibinfo {year} {2015}{\natexlab{a}})},\ \Eprint
  {http://arxiv.org/abs/1407.1236} {arXiv:1407.1236} \BibitemShut {NoStop}%
\bibitem [{\citenamefont {{Zhao}}\ \emph
  {et~al.}(2015{\natexlab{b}})\citenamefont {{Zhao}}, \citenamefont
  {{Kitaura}}, \citenamefont {{Chuang}}, \citenamefont {{Prada}}, \citenamefont
  {{Yepes}},\ and\ \citenamefont {{Tao}}}]{Zhao15}%
  \BibitemOpen
  \bibfield  {author} {\bibinfo {author} {\bibfnamefont {C.}~\bibnamefont
  {{Zhao}}}, \bibinfo {author} {\bibfnamefont {F.-S.}\ \bibnamefont
  {{Kitaura}}}, \bibinfo {author} {\bibfnamefont {C.-H.}\ \bibnamefont
  {{Chuang}}}, \bibinfo {author} {\bibfnamefont {F.}~\bibnamefont {{Prada}}},
  \bibinfo {author} {\bibfnamefont {G.}~\bibnamefont {{Yepes}}}, \ and\
  \bibinfo {author} {\bibfnamefont {C.}~\bibnamefont {{Tao}}},\ }\href
  {\doibase 10.1093/mnras/stv1262} {\bibfield  {journal} {\bibinfo  {journal}
  {\mnras}\ }\textbf {\bibinfo {volume} {451}},\ \bibinfo {pages} {4266}
  (\bibinfo {year} {2015}{\natexlab{b}})},\ \Eprint
  {http://arxiv.org/abs/1501.05520} {arXiv:1501.05520} \BibitemShut {NoStop}%
\bibitem [{\citenamefont {{Chuang}}\ \emph {et~al.}(2015)\citenamefont
  {{Chuang}}, \citenamefont {{Zhao}}, \citenamefont {{Prada}}, \citenamefont
  {{Munari}}, \citenamefont {{Avila}}, \citenamefont {{Izard}}, \citenamefont
  {{Kitaura}}, \citenamefont {{Manera}},\ and\ \citenamefont
  {et~al.}}]{ChuangComp15}%
  \BibitemOpen
  \bibfield  {author} {\bibinfo {author} {\bibfnamefont {C.-H.}\ \bibnamefont
  {{Chuang}}}, \bibinfo {author} {\bibfnamefont {C.}~\bibnamefont {{Zhao}}},
  \bibinfo {author} {\bibfnamefont {F.}~\bibnamefont {{Prada}}}, \bibinfo
  {author} {\bibfnamefont {E.}~\bibnamefont {{Munari}}}, \bibinfo {author}
  {\bibfnamefont {S.}~\bibnamefont {{Avila}}}, \bibinfo {author} {\bibfnamefont
  {A.}~\bibnamefont {{Izard}}}, \bibinfo {author} {\bibfnamefont {F.-S.}\
  \bibnamefont {{Kitaura}}}, \bibinfo {author} {\bibfnamefont {M.}~\bibnamefont
  {{Manera}}}, \ and\ \bibinfo {author} {\bibnamefont {et~al.}},\ }\href
  {\doibase 10.1093/mnras/stv1289} {\bibfield  {journal} {\bibinfo  {journal}
  {\mnras}\ }\textbf {\bibinfo {volume} {452}},\ \bibinfo {pages} {686}
  (\bibinfo {year} {2015})},\ \Eprint {http://arxiv.org/abs/1412.7729}
  {arXiv:1412.7729} \BibitemShut {NoStop}%
\bibitem [{\citenamefont {{Wertheim}}(1963)}]{W63}%
  \BibitemOpen
  \bibfield  {author} {\bibinfo {author} {\bibfnamefont {M.~S.}\ \bibnamefont
  {{Wertheim}}},\ }\href {\doibase 10.1103/PhysRevLett.10.321} {\bibfield
  {journal} {\bibinfo  {journal} {Physical Review Letters}\ }\textbf {\bibinfo
  {volume} {10}},\ \bibinfo {pages} {321} (\bibinfo {year} {1963})}\BibitemShut
  {NoStop}%
\bibitem [{\citenamefont {{Alam et al.}}(2015)}]{Alam15}%
  \BibitemOpen
  \bibfield  {author} {\bibinfo {author} {\bibfnamefont {S.}~\bibnamefont
  {{Alam et al.}}},\ }\href {\doibase 10.1088/0067-0049/219/1/12} {\bibfield
  {journal} {\bibinfo  {journal} {\apjs}\ }\textbf {\bibinfo {volume} {219}},\
  \bibinfo {eid} {12} (\bibinfo {year} {2015})},\ \Eprint
  {http://arxiv.org/abs/1501.00963} {arXiv:1501.00963 [astro-ph.IM]}
  \BibitemShut {NoStop}%
\bibitem [{\citenamefont {{Eisenstein}}\ \emph {et~al.}(2011)\citenamefont
  {{Eisenstein}}, \citenamefont {{Weinberg}}, \citenamefont {{Agol}},
  \citenamefont {{Aihara}}, \citenamefont {{Allende Prieto}}, \citenamefont
  {{Anderson}}, \citenamefont {{Arns}}, \citenamefont {{Aubourg}},\ and\
  \citenamefont {et~al.}}]{2011AJ....142...72E}%
  \BibitemOpen
  \bibfield  {author} {\bibinfo {author} {\bibfnamefont {D.~J.}\ \bibnamefont
  {{Eisenstein}}}, \bibinfo {author} {\bibfnamefont {D.~H.}\ \bibnamefont
  {{Weinberg}}}, \bibinfo {author} {\bibfnamefont {E.}~\bibnamefont {{Agol}}},
  \bibinfo {author} {\bibfnamefont {H.}~\bibnamefont {{Aihara}}}, \bibinfo
  {author} {\bibfnamefont {C.}~\bibnamefont {{Allende Prieto}}}, \bibinfo
  {author} {\bibfnamefont {S.~F.}\ \bibnamefont {{Anderson}}}, \bibinfo
  {author} {\bibfnamefont {J.~A.}\ \bibnamefont {{Arns}}}, \bibinfo {author}
  {\bibfnamefont {{\'E}.}~\bibnamefont {{Aubourg}}}, \ and\ \bibinfo {author}
  {\bibnamefont {et~al.}},\ }\href {\doibase 10.1088/0004-6256/142/3/72}
  {\bibfield  {journal} {\bibinfo  {journal} {\aj}\ }\textbf {\bibinfo {volume}
  {142}},\ \bibinfo {eid} {72} (\bibinfo {year} {2011})},\ \Eprint
  {http://arxiv.org/abs/1101.1529} {arXiv:1101.1529 [astro-ph.IM]} \BibitemShut
  {NoStop}%
\bibitem [{\citenamefont {{Gunn}}\ \emph {et~al.}(2006)\citenamefont {{Gunn}},
  \citenamefont {{Siegmund}}, \citenamefont {{Mannery}}, \citenamefont
  {{Owen}}, \citenamefont {{Hull}}, \citenamefont {{Leger}}, \citenamefont
  {{Carey}}, \citenamefont {{Knapp}},\ and\ \citenamefont
  {et~al.}}]{2006AJ....131.2332G}%
  \BibitemOpen
  \bibfield  {author} {\bibinfo {author} {\bibfnamefont {J.~E.}\ \bibnamefont
  {{Gunn}}}, \bibinfo {author} {\bibfnamefont {W.~A.}\ \bibnamefont
  {{Siegmund}}}, \bibinfo {author} {\bibfnamefont {E.~J.}\ \bibnamefont
  {{Mannery}}}, \bibinfo {author} {\bibfnamefont {R.~E.}\ \bibnamefont
  {{Owen}}}, \bibinfo {author} {\bibfnamefont {C.~L.}\ \bibnamefont {{Hull}}},
  \bibinfo {author} {\bibfnamefont {R.~F.}\ \bibnamefont {{Leger}}}, \bibinfo
  {author} {\bibfnamefont {L.~N.}\ \bibnamefont {{Carey}}}, \bibinfo {author}
  {\bibfnamefont {G.~R.}\ \bibnamefont {{Knapp}}}, \ and\ \bibinfo {author}
  {\bibnamefont {et~al.}},\ }\href {\doibase 10.1086/500975} {\bibfield
  {journal} {\bibinfo  {journal} {\aj}\ }\textbf {\bibinfo {volume} {131}},\
  \bibinfo {pages} {2332} (\bibinfo {year} {2006})},\ \Eprint
  {http://arxiv.org/abs/astro-ph/0602326} {astro-ph/0602326} \BibitemShut
  {NoStop}%
\bibitem [{\citenamefont {{Smee}}\ \emph {et~al.}(2013)\citenamefont {{Smee}},
  \citenamefont {{Gunn}}, \citenamefont {{Uomoto}}, \citenamefont {{Roe}},
  \citenamefont {{Schlegel}}, \citenamefont {{Rockosi}}, \citenamefont
  {{Carr}}, \citenamefont {{Leger}},\ and\ \citenamefont
  {et~al.}}]{2013AJ....146...32S}%
  \BibitemOpen
  \bibfield  {author} {\bibinfo {author} {\bibfnamefont {S.~A.}\ \bibnamefont
  {{Smee}}}, \bibinfo {author} {\bibfnamefont {J.~E.}\ \bibnamefont {{Gunn}}},
  \bibinfo {author} {\bibfnamefont {A.}~\bibnamefont {{Uomoto}}}, \bibinfo
  {author} {\bibfnamefont {N.}~\bibnamefont {{Roe}}}, \bibinfo {author}
  {\bibfnamefont {D.}~\bibnamefont {{Schlegel}}}, \bibinfo {author}
  {\bibfnamefont {C.~M.}\ \bibnamefont {{Rockosi}}}, \bibinfo {author}
  {\bibfnamefont {M.~A.}\ \bibnamefont {{Carr}}}, \bibinfo {author}
  {\bibfnamefont {F.}~\bibnamefont {{Leger}}}, \ and\ \bibinfo {author}
  {\bibnamefont {et~al.}},\ }\href {\doibase 10.1088/0004-6256/146/2/32}
  {\bibfield  {journal} {\bibinfo  {journal} {\aj}\ }\textbf {\bibinfo {volume}
  {146}},\ \bibinfo {eid} {32} (\bibinfo {year} {2013})},\ \Eprint
  {http://arxiv.org/abs/1208.2233} {arXiv:1208.2233 [astro-ph.IM]} \BibitemShut
  {NoStop}%
\bibitem [{\citenamefont {{Bolton}}\ \emph {et~al.}(2012)\citenamefont
  {{Bolton}}, \citenamefont {{Schlegel}}, \citenamefont {{Aubourg}},
  \citenamefont {{Bailey}}, \citenamefont {{Bhardwaj}}, \citenamefont
  {{Brownstein}}, \citenamefont {{Burles}}, \citenamefont {{Chen}},\ and\
  \citenamefont {et~al.}}]{2012AJ....144..144B}%
  \BibitemOpen
  \bibfield  {author} {\bibinfo {author} {\bibfnamefont {A.~S.}\ \bibnamefont
  {{Bolton}}}, \bibinfo {author} {\bibfnamefont {D.~J.}\ \bibnamefont
  {{Schlegel}}}, \bibinfo {author} {\bibfnamefont {{\'E}.}~\bibnamefont
  {{Aubourg}}}, \bibinfo {author} {\bibfnamefont {S.}~\bibnamefont {{Bailey}}},
  \bibinfo {author} {\bibfnamefont {V.}~\bibnamefont {{Bhardwaj}}}, \bibinfo
  {author} {\bibfnamefont {J.~R.}\ \bibnamefont {{Brownstein}}}, \bibinfo
  {author} {\bibfnamefont {S.}~\bibnamefont {{Burles}}}, \bibinfo {author}
  {\bibfnamefont {Y.-M.}\ \bibnamefont {{Chen}}}, \ and\ \bibinfo {author}
  {\bibnamefont {et~al.}},\ }\href {\doibase 10.1088/0004-6256/144/5/144}
  {\bibfield  {journal} {\bibinfo  {journal} {\aj}\ }\textbf {\bibinfo {volume}
  {144}},\ \bibinfo {eid} {144} (\bibinfo {year} {2012})},\ \Eprint
  {http://arxiv.org/abs/1207.7326} {arXiv:1207.7326} \BibitemShut {NoStop}%
\bibitem [{\citenamefont {{Reid}}\ \emph {et~al.}(2015)\citenamefont {{Reid}},
  \citenamefont {{Ho}}, \citenamefont {{Padmanabhan}}, \citenamefont
  {{Percival}}, \citenamefont {{Tinker}}, \citenamefont {{Tojeiro}},
  \citenamefont {{White}}, \citenamefont {{Eisenstein}},\ and\ \citenamefont
  {et~al.}}]{2015arXiv150906529R}%
  \BibitemOpen
  \bibfield  {author} {\bibinfo {author} {\bibfnamefont {B.}~\bibnamefont
  {{Reid}}}, \bibinfo {author} {\bibfnamefont {S.}~\bibnamefont {{Ho}}},
  \bibinfo {author} {\bibfnamefont {N.}~\bibnamefont {{Padmanabhan}}}, \bibinfo
  {author} {\bibfnamefont {W.~J.}\ \bibnamefont {{Percival}}}, \bibinfo
  {author} {\bibfnamefont {J.}~\bibnamefont {{Tinker}}}, \bibinfo {author}
  {\bibfnamefont {R.}~\bibnamefont {{Tojeiro}}}, \bibinfo {author}
  {\bibfnamefont {M.}~\bibnamefont {{White}}}, \bibinfo {author} {\bibfnamefont
  {D.~J.}\ \bibnamefont {{Eisenstein}}}, \ and\ \bibinfo {author} {\bibnamefont
  {et~al.}},\ }\href@noop {} {\bibfield  {journal} {\bibinfo  {journal} {ArXiv
  e-prints}\ } (\bibinfo {year} {2015})},\ \Eprint
  {http://arxiv.org/abs/1509.06529} {arXiv:1509.06529} \BibitemShut {NoStop}%
\bibitem [{\citenamefont {{Kitaura}}\ \emph
  {et~al.}(2015{\natexlab{b}})\citenamefont {{Kitaura}}, \citenamefont
  {{Rodr{\'{\i}}guez-Torres}}, \citenamefont {{Chuang}}, \citenamefont
  {{Zhao}}, \citenamefont {{Prada}}, \citenamefont {{Gil-Marin}}, \citenamefont
  {{Guo}}, \citenamefont {{Yepes}},\ and\ \citenamefont
  {et~al.}}]{2015arXiv150906400K}%
  \BibitemOpen
  \bibfield  {author} {\bibinfo {author} {\bibfnamefont {F.-S.}\ \bibnamefont
  {{Kitaura}}}, \bibinfo {author} {\bibfnamefont {S.}~\bibnamefont
  {{Rodr{\'{\i}}guez-Torres}}}, \bibinfo {author} {\bibfnamefont {C.-H.}\
  \bibnamefont {{Chuang}}}, \bibinfo {author} {\bibfnamefont {C.}~\bibnamefont
  {{Zhao}}}, \bibinfo {author} {\bibfnamefont {F.}~\bibnamefont {{Prada}}},
  \bibinfo {author} {\bibfnamefont {H.}~\bibnamefont {{Gil-Marin}}}, \bibinfo
  {author} {\bibfnamefont {H.}~\bibnamefont {{Guo}}}, \bibinfo {author}
  {\bibfnamefont {G.}~\bibnamefont {{Yepes}}}, \ and\ \bibinfo {author}
  {\bibnamefont {et~al.}},\ }\href@noop {} {\bibfield  {journal} {\bibinfo
  {journal} {ArXiv e-prints}\ } (\bibinfo {year} {2015}{\natexlab{b}})},\
  \Eprint {http://arxiv.org/abs/1509.06400} {arXiv:1509.06400} \BibitemShut
  {NoStop}%
\bibitem [{\citenamefont {{Rodr{\'{\i}}guez-Torres}}\ \emph
  {et~al.}(2015)\citenamefont {{Rodr{\'{\i}}guez-Torres}}, \citenamefont
  {{Prada}}, \citenamefont {{Chuang}}, \citenamefont {{Guo}}, \citenamefont
  {{Klypin}}, \citenamefont {{Behroozi}}, \citenamefont {{Hahn}}, \citenamefont
  {{Comparat}},\ and\ \citenamefont {et~al.}}]{2015arXiv150906404R}%
  \BibitemOpen
  \bibfield  {author} {\bibinfo {author} {\bibfnamefont {S.~A.}\ \bibnamefont
  {{Rodr{\'{\i}}guez-Torres}}}, \bibinfo {author} {\bibfnamefont
  {F.}~\bibnamefont {{Prada}}}, \bibinfo {author} {\bibfnamefont {C.-H.}\
  \bibnamefont {{Chuang}}}, \bibinfo {author} {\bibfnamefont {H.}~\bibnamefont
  {{Guo}}}, \bibinfo {author} {\bibfnamefont {A.}~\bibnamefont {{Klypin}}},
  \bibinfo {author} {\bibfnamefont {P.}~\bibnamefont {{Behroozi}}}, \bibinfo
  {author} {\bibfnamefont {C.~H.}\ \bibnamefont {{Hahn}}}, \bibinfo {author}
  {\bibfnamefont {J.}~\bibnamefont {{Comparat}}}, \ and\ \bibinfo {author}
  {\bibnamefont {et~al.}},\ }\href@noop {} {\bibfield  {journal} {\bibinfo
  {journal} {ArXiv e-prints}\ } (\bibinfo {year} {2015})},\ \Eprint
  {http://arxiv.org/abs/1509.06404} {arXiv:1509.06404} \BibitemShut {NoStop}%
\bibitem [{Note1()}]{Note1}%
  \BibitemOpen
  \bibinfo {note} {\protect \burl
  {http://data.sdss.org/datamodel/files/BOSS_LSS_REDUX/dr11_patchy_mocks/}}\BibitemShut
  {NoStop}%
\bibitem [{\citenamefont {{Chuang}}\ \emph {et~al.}(2012)\citenamefont
  {{Chuang}}, \citenamefont {{Wang}},\ and\ \citenamefont
  {{Hemantha}}}]{Chuang12}%
  \BibitemOpen
  \bibfield  {author} {\bibinfo {author} {\bibfnamefont {C.-H.}\ \bibnamefont
  {{Chuang}}}, \bibinfo {author} {\bibfnamefont {Y.}~\bibnamefont {{Wang}}}, \
  and\ \bibinfo {author} {\bibfnamefont {M.~D.~P.}\ \bibnamefont
  {{Hemantha}}},\ }\href {\doibase 10.1111/j.1365-2966.2012.20971.x} {\bibfield
   {journal} {\bibinfo  {journal} {\mnras}\ }\textbf {\bibinfo {volume}
  {423}},\ \bibinfo {pages} {1474} (\bibinfo {year} {2012})},\ \Eprint
  {http://arxiv.org/abs/1008.4822} {arXiv:1008.4822 [astro-ph.CO]} \BibitemShut
  {NoStop}%
\bibitem [{\citenamefont {{Ross}}\ \emph {et~al.}(2012)\citenamefont {{Ross}},
  \citenamefont {{Percival}}, \citenamefont {{S{\'a}nchez}}, \citenamefont
  {{Samushia}}, \citenamefont {{Ho}}, \citenamefont {{Kazin}}, \citenamefont
  {{Manera}}, \citenamefont {{Reid}},\ and\ \citenamefont {et~al.}}]{Ross2012}%
  \BibitemOpen
  \bibfield  {author} {\bibinfo {author} {\bibfnamefont {A.~J.}\ \bibnamefont
  {{Ross}}}, \bibinfo {author} {\bibfnamefont {W.~J.}\ \bibnamefont
  {{Percival}}}, \bibinfo {author} {\bibfnamefont {A.~G.}\ \bibnamefont
  {{S{\'a}nchez}}}, \bibinfo {author} {\bibfnamefont {L.}~\bibnamefont
  {{Samushia}}}, \bibinfo {author} {\bibfnamefont {S.}~\bibnamefont {{Ho}}},
  \bibinfo {author} {\bibfnamefont {E.}~\bibnamefont {{Kazin}}}, \bibinfo
  {author} {\bibfnamefont {M.}~\bibnamefont {{Manera}}}, \bibinfo {author}
  {\bibfnamefont {B.}~\bibnamefont {{Reid}}}, \ and\ \bibinfo {author}
  {\bibnamefont {et~al.}},\ }\href {\doibase 10.1111/j.1365-2966.2012.21235.x}
  {\bibfield  {journal} {\bibinfo  {journal} {\mnras}\ }\textbf {\bibinfo
  {volume} {424}},\ \bibinfo {pages} {564} (\bibinfo {year} {2012})},\ \Eprint
  {http://arxiv.org/abs/1203.6499} {arXiv:1203.6499} \BibitemShut {NoStop}%
\bibitem [{\citenamefont {{Chuang}}\ \emph {et~al.}(2013)\citenamefont
  {{Chuang}}, \citenamefont {{Prada}}, \citenamefont {{Beutler}}, \citenamefont
  {{Eisenstein}}, \citenamefont {{Escoffier}}, \citenamefont {{Ho}},
  \citenamefont {{Kneib}}, \citenamefont {{Manera}},\ and\ \citenamefont
  {et~al.}}]{Chuang2013d}%
  \BibitemOpen
  \bibfield  {author} {\bibinfo {author} {\bibfnamefont {C.-H.}\ \bibnamefont
  {{Chuang}}}, \bibinfo {author} {\bibfnamefont {F.}~\bibnamefont {{Prada}}},
  \bibinfo {author} {\bibfnamefont {F.}~\bibnamefont {{Beutler}}}, \bibinfo
  {author} {\bibfnamefont {D.~J.}\ \bibnamefont {{Eisenstein}}}, \bibinfo
  {author} {\bibfnamefont {S.}~\bibnamefont {{Escoffier}}}, \bibinfo {author}
  {\bibfnamefont {S.}~\bibnamefont {{Ho}}}, \bibinfo {author} {\bibfnamefont
  {J.-P.}\ \bibnamefont {{Kneib}}}, \bibinfo {author} {\bibfnamefont
  {M.}~\bibnamefont {{Manera}}}, \ and\ \bibinfo {author} {\bibnamefont
  {et~al.}},\ }\href@noop {} {\bibfield  {journal} {\bibinfo  {journal} {ArXiv
  e-prints}\ } (\bibinfo {year} {2013})},\ \Eprint
  {http://arxiv.org/abs/1312.4889} {arXiv:1312.4889 [astro-ph.CO]} \BibitemShut
  {NoStop}%
\bibitem [{\citenamefont {{Frieman}}\ and\ \citenamefont
  {{Gaztanaga}}(1994)}]{Frieman1994}%
  \BibitemOpen
  \bibfield  {author} {\bibinfo {author} {\bibfnamefont {J.~A.}\ \bibnamefont
  {{Frieman}}}\ and\ \bibinfo {author} {\bibfnamefont {E.}~\bibnamefont
  {{Gaztanaga}}},\ }\href {\doibase 10.1086/173995} {\bibfield  {journal}
  {\bibinfo  {journal} {\apj}\ }\textbf {\bibinfo {volume} {425}},\ \bibinfo
  {pages} {392} (\bibinfo {year} {1994})},\ \Eprint
  {http://arxiv.org/abs/astro-ph/9306018} {astro-ph/9306018} \BibitemShut
  {NoStop}%
\bibitem [{\citenamefont {{Schmittfull}}\ \emph {et~al.}(2015)\citenamefont
  {{Schmittfull}}, \citenamefont {{Feng}}, \citenamefont {{Beutler}},
  \citenamefont {{Sherwin}},\ and\ \citenamefont
  {{Chu}}}]{2015PhRvD..92l3522S}%
  \BibitemOpen
  \bibfield  {author} {\bibinfo {author} {\bibfnamefont {M.}~\bibnamefont
  {{Schmittfull}}}, \bibinfo {author} {\bibfnamefont {Y.}~\bibnamefont
  {{Feng}}}, \bibinfo {author} {\bibfnamefont {F.}~\bibnamefont {{Beutler}}},
  \bibinfo {author} {\bibfnamefont {B.}~\bibnamefont {{Sherwin}}}, \ and\
  \bibinfo {author} {\bibfnamefont {M.~Y.}\ \bibnamefont {{Chu}}},\ }\href
  {\doibase 10.1103/PhysRevD.92.123522} {\bibfield  {journal} {\bibinfo
  {journal} {\prd}\ }\textbf {\bibinfo {volume} {92}},\ \bibinfo {eid} {123522}
  (\bibinfo {year} {2015})},\ \Eprint {http://arxiv.org/abs/1508.06972}
  {arXiv:1508.06972} \BibitemShut {NoStop}%
\bibitem [{\citenamefont {{Achitouv}}\ and\ \citenamefont
  {{Blake}}(2015)}]{2015PhRvD..92h3523A}%
  \BibitemOpen
  \bibfield  {author} {\bibinfo {author} {\bibfnamefont {I.}~\bibnamefont
  {{Achitouv}}}\ and\ \bibinfo {author} {\bibfnamefont {C.}~\bibnamefont
  {{Blake}}},\ }\href {\doibase 10.1103/PhysRevD.92.083523} {\bibfield
  {journal} {\bibinfo  {journal} {\prd}\ }\textbf {\bibinfo {volume} {92}},\
  \bibinfo {eid} {083523} (\bibinfo {year} {2015})},\ \Eprint
  {http://arxiv.org/abs/1507.03584} {arXiv:1507.03584} \BibitemShut {NoStop}%
\bibitem [{\citenamefont {{Slepian}}\ \emph {et~al.}(2015)\citenamefont
  {{Slepian}}, \citenamefont {{Eisenstein}}, \citenamefont {{Beutler}},
  \citenamefont {{Cuesta}}, \citenamefont {{Ge}}, \citenamefont
  {{Gil-Mar{\'{\i}}n}}, \citenamefont {{Ho}}, \citenamefont {{Kitaura}},
  \citenamefont {{McBride}}, \citenamefont {{Nichol}}, \citenamefont
  {{Percival}}, \citenamefont {{Rodr{\'{\i}}guez-Torres}}, \citenamefont
  {{Ross}}, \citenamefont {{Scoccimarro}}, \citenamefont {{Seo}}, \citenamefont
  {{Tinker}}, \citenamefont {{Tojeiro}},\ and\ \citenamefont
  {{Vargas-Maga{\~n}a}}}]{2015arXiv151202231S}%
  \BibitemOpen
  \bibfield  {author} {\bibinfo {author} {\bibfnamefont {Z.}~\bibnamefont
  {{Slepian}}}, \bibinfo {author} {\bibfnamefont {D.~J.}\ \bibnamefont
  {{Eisenstein}}}, \bibinfo {author} {\bibfnamefont {F.}~\bibnamefont
  {{Beutler}}}, \bibinfo {author} {\bibfnamefont {A.~J.}\ \bibnamefont
  {{Cuesta}}}, \bibinfo {author} {\bibfnamefont {J.}~\bibnamefont {{Ge}}},
  \bibinfo {author} {\bibfnamefont {H.}~\bibnamefont {{Gil-Mar{\'{\i}}n}}},
  \bibinfo {author} {\bibfnamefont {S.}~\bibnamefont {{Ho}}}, \bibinfo {author}
  {\bibfnamefont {F.-S.}\ \bibnamefont {{Kitaura}}}, \bibinfo {author}
  {\bibfnamefont {C.~K.}\ \bibnamefont {{McBride}}}, \bibinfo {author}
  {\bibfnamefont {R.~C.}\ \bibnamefont {{Nichol}}}, \bibinfo {author}
  {\bibfnamefont {W.~J.}\ \bibnamefont {{Percival}}}, \bibinfo {author}
  {\bibfnamefont {S.}~\bibnamefont {{Rodr{\'{\i}}guez-Torres}}}, \bibinfo
  {author} {\bibfnamefont {A.~J.}\ \bibnamefont {{Ross}}}, \bibinfo {author}
  {\bibfnamefont {R.}~\bibnamefont {{Scoccimarro}}}, \bibinfo {author}
  {\bibfnamefont {H.-J.}\ \bibnamefont {{Seo}}}, \bibinfo {author}
  {\bibfnamefont {J.}~\bibnamefont {{Tinker}}}, \bibinfo {author}
  {\bibfnamefont {R.}~\bibnamefont {{Tojeiro}}}, \ and\ \bibinfo {author}
  {\bibfnamefont {M.}~\bibnamefont {{Vargas-Maga{\~n}a}}},\ }\href@noop {}
  {\bibfield  {journal} {\bibinfo  {journal} {ArXiv e-prints}\ } (\bibinfo
  {year} {2015})},\ \Eprint {http://arxiv.org/abs/1512.02231}
  {arXiv:1512.02231} \BibitemShut {NoStop}%
\bibitem [{\citenamefont {{Eisenstein}}\ \emph {et~al.}(2007)\citenamefont
  {{Eisenstein}}, \citenamefont {{Seo}}, \citenamefont {{Sirko}},\ and\
  \citenamefont {{Spergel}}}]{ESS07}%
  \BibitemOpen
  \bibfield  {author} {\bibinfo {author} {\bibfnamefont {D.~J.}\ \bibnamefont
  {{Eisenstein}}}, \bibinfo {author} {\bibfnamefont {H.-J.}\ \bibnamefont
  {{Seo}}}, \bibinfo {author} {\bibfnamefont {E.}~\bibnamefont {{Sirko}}}, \
  and\ \bibinfo {author} {\bibfnamefont {D.~N.}\ \bibnamefont {{Spergel}}},\
  }\href {\doibase 10.1086/518712} {\bibfield  {journal} {\bibinfo  {journal}
  {\apj}\ }\textbf {\bibinfo {volume} {664}},\ \bibinfo {pages} {675} (\bibinfo
  {year} {2007})},\ \Eprint {http://arxiv.org/abs/astro-ph/0604362}
  {astro-ph/0604362} \BibitemShut {NoStop}%
\end{thebibliography}%
